\documentclass[11pt]{article}

\usepackage{amsmath}
\usepackage{graphicx}
\usepackage{indentfirst}
\usepackage{amssymb}
\usepackage{cite}
\usepackage{color}
\usepackage{subfigure}
\usepackage{xcolor}
\usepackage[breaklinks=true]{hyperref}

\begin{document}

\title{Observational appearance and additional photon rings of the asymmetric thin-shell wormhole in Horndeski theory}

\date{}
\maketitle

\begin{center}
	Zhi Luo$~^{a}$,~Hao Yu$~^{a,b}$,~Jin Li~$^{a,b}$\footnote{cqstarv@hotmail.com, the corresponding author} \\
\end{center}

\begin{center}
	$^a$ Department of Physics, Chongqing University, Chongqing 401331, China\\
	$^b$  Department of Physics and Chongqing Key Laboratory for Strongly Coupled Physics, Chongqing University, Chongqing 401331, China
\end{center}

\vskip 0.6in
{\abstract 
{
{In this paper, we study the observational appearance of the asymmetric thin-shell wormhole (ATW) in Horndeski theory by employing the ray-tracing method. We first calculate the effective potential and null geodesic of the ATW, and then we obtain the deflection angle of the photon in the ATW spacetime. Based on the impact parameter of the photon, the trajectory of the photon can be classified into three cases. Two typical emission models of the thin accretion disk are considered to analyze the observational appearance of the ATW. By comparing the observational appearances of the ATW and a black hole with the same mass parameter, we find additional features in the observational appearance of the ATW, such as the ``lensing band" and ``photon ring group".}
}


\thispagestyle{empty}
\newpage
\setcounter{page}{1}

\section{Introduction}\label{sec1}
The Laser Interferometer Gravitational-Wave Observatory (LIGO) made its first detection of a gravitational wave signal in 2015, which was generated by the merger of two black holes (BHs) with stellar masses~\cite{LIGOScientific:2016sjg,LIGOScientific:2016vlm,LIGOScientific:2016aoc}. This gravitational wave event provides indirect evidence bolstering the existence of BHs. Recently, the Event Horizon Telescope (EHT) unveiled the optical images of $\mathrm{M} 87^*$~\cite{EventHorizonTelescope:2019dse,EventHorizonTelescope:2019uob,EventHorizonTelescope:2019jan,EventHorizonTelescope:2019ths,EventHorizonTelescope:2019pgp,EventHorizonTelescope:2019ggy} and Sagittarius~$\mathrm{A} ^*$~\cite{EventHorizonTelescope:2022xnr,EventHorizonTelescope:2022vjs,EventHorizonTelescope:2022wok,EventHorizonTelescope:2022exc,EventHorizonTelescope:2022urf,EventHorizonTelescope:2022xqj}, which provide compelling evidence for the existence of supermassive BHs at the centers of galaxies. These images depict dark regions surrounded by luminous rings of BHs, which are formed through the interactions between photons and the photon spheres (i.e. the critical curve) of the BHs~\cite{Gralla:2019xty}. The central region enclosed by the critical curve of a BH is commonly referred to as the BH shadow~\cite{Perlick:2021aok}. The shadow image of the BH offers valuable insights into the behavior of jets and the surrounding matter, enabling us to determine important physical characteristics such as its mass, rotation, and charge. Synge and Luminet were the first to propose an equation utilizing a critical impact parameter to determine the angular radius of the photon capture region for a Schwarzschild BH~\cite{Synge:1966okc,Luminet:1979nyg}. Bardeen presented the first depiction of a rotating Kerr BH shadow, illustrating how its spin leads to deformations in its silhouette caused by the dragging effect~\cite{Bardeen:1973tla,Hioki:2008zw,Eiroa:2017uuq}.

Although the findings of EHT lie within the framework of General Relativity (GR), they also open up the possibility of ultracompact objects (UCOs) or BHs in gravity theories beyond GR. In recent years, there has been extensive research investigating the shadows of BHs within the context of various modified theories of gravity~\cite{Amarilla:2010zq,Amarilla:2011fx,Amarilla:2013sj,Amir:2017slq,Singh:2017vfr,Mizuno:2018lxz,Vagnozzi:2019apd,Banerjee:2019nnj,Bambi:2008jg,Bambi:2010hf,Atamurotov:2013sca,Papnoi:2014aaa,Atamurotov:2015nra,Wang:2017qhh,Guo:2018kis,Yan:2019etp,Konoplya:2019sns,Bambi:2019tjh,Allahyari:2019jqz,Vagnozzi:2020quf,Khodadi:2020jij,Vagnozzi:2022moj,Tsukamoto:2014tja,Tsukamoto:2017fxq,Hu:2020usx,Zhong:2021mty,Peng:2020wun,Hou:2022eev,Hou:2021okc}. For a comprehensive overview of these BH shadows, one can refer to Ref.~\cite{Cunha:2018acu}. 

It is worth noting that the UCOs located at the centers of galaxies cannot definitively be classified as BHs based on the photon spheres, because in addition to BHs, there are also other UCOs that possess similar photon spheres~\cite{Cunha:2017qtt,Guo:2020qwk}. For example, certain UCOs can produce an effective shadow that mimics the BH shadow~\cite{Rosa:2023hfm,Tamm:2023wvn,Abdikamalov:2019ztb,Narzilloev:2020peq,Herdeiro:2021lwl,Wang:2023nwd}. Therefore, it is crucial to find a way to distinguish between BHs and other UCOs such as boson stars~\cite{Cunha:2016bjh,Rosa:2022tfv,Rosa:2022toh,Rosa:2023qcv}, wormholes~\cite{Nedkova:2013msa,Bugaev:2021dna,Kasuya:2021cpk,Olmo:2023lil,Bronnikov:2021liv,Tsukamoto:2021caq,Tsukamoto:2022vkt} and so forth. Recently, Wang \textit{et al.}~\cite{Wang:2020emr} studied the shadow of an asymmetric thin-shell wormhole (ATW) and found that the ATW shadow is always smaller than that of a BH with the same parameter settings. The discovery provides a method to distinguish between BHs and ATWs through the observations of EHT. Subsequently, the double shadows of ATWs were studied in Refs.~\cite{Wielgus:2020uqz,Guerrero:2021pxt,Tsukamoto:2021fpp}. Peng \textit{et al.}~\cite{Peng:2021osd} investigated the observational appearance of the emission disk surrounding a Schwarzschild ATW and found additional photon rings and an extra lensing band. These findings also serve as observational signatures that can be used to distinguishing between ATWs and BHs. Furthermore, Ref.~\cite{Guo:2022iiy} examined the observational appearance of an ATW in the Hayward spacetime, while Ref.~\cite{Chen:2022tog} investigated the observational appearance of a freely-falling star in an ATW. Inspired by these studies, in this paper we explore the observational appearance of the ATW surrounded by a thin accretion disk in Horndeski theory.

In 1974, Horndeski constructed the most general scalar-tensor theory with second-order derivative equations of motion~\cite{Horndeski:1974wa}. In Horndeski theory, the action is expressed as
\begin{equation}\label{eq1}
	S=\int d^4 x \sqrt{-g}\left(\mathcal{L}_2+\mathcal{L}_3+\mathcal{L}_4+\mathcal{L}_5\right),
\end{equation}
where
\begin{equation}
	\begin{aligned}
		\mathcal{L}_2=  G_2(X), \quad \mathcal{L}_3=-G_3(X) \square \phi, \quad \mathcal{L}_4=  G_4(X) \mathcal{R}+G_{4 X}\left[(\square \phi)^2-\left(\nabla_\mu \nabla_\nu \phi\right)^2\right], \\
		\mathcal{L}_5=  G_5(X) G_{\mu \nu} \nabla^\mu \nabla^\nu \phi-\frac{G_{5 X}}{6}\left[(\square \phi)^3-3 \square \phi\left(\nabla_\mu \nabla_\nu \phi\right)^2+2\left(\nabla_\mu \nabla_\nu \phi\right)^3\right].
	\end{aligned}
\end{equation}
Here, $\mathcal{R}$ is the Ricci scalar, $G_{\mu \nu}$ is the Einstein tensor, and $G_{i}=G_{i}(\phi, X)$ denotes the function of the scalar field $\phi$ and its kinetic energy $X=$ $-\partial_\mu \phi \partial^\mu \phi / 2$. The subscript $X$ represents the derivative with respect to $X$ (i.e., $G_{iX}=dG_{i}(\phi, X)/dX$). As a prominent scalar-tensor theory, Horndeski theory contains various gravity theories, such as GR, quintessence, $k$-essence, covariant Galileon, and others~\cite{Deffayet:2009wt,Deffayet:2011gz,Charmousis:2011bf}. In 2015, Gleyzes~\textit{et al.} introduced a new class of scalar-tensor theories to extend Horndeski theory (known as ``beyond Horndeski theory''), thereby avoiding the effects of Ostrogradski instabilities~\cite{Gleyzes:2014dya,Gleyzes:2014qga}. Then, Babichev~\textit{et al.} found spherically symmetric and static BH solutions in shift-symmetric Horndeski and beyond Horndeski theory~\cite{Babichev:2017guv}. In recent years, there has been a significant attention given to the investigation of BHs in (beyond) Horndeski theory~\cite{Rinaldi:2012vy,Cisterna:2014nua,Anabalon:2013oea,Babichev:2023psy,Charmousis:2014zaa,Minamitsuji:2013ura,Babichev:2016rlq,Babichev:2017guv,Feng:2015oea,Cvetic:2016bxi}. This study aims to investigate the observational appearances of ATWs and BHs in Horndeski theory. By conducting the study, we can potentially differentiate between ATWs and BHs in Horndeski theory, while also examining the impact of parameters in Horndeski theory on the results.

This paper is organized as follows. In Sec.~\ref{sec:2}~, we construct the ATW in Horndeski theory and  study its effective potential and geodesic. We introduce the light deflection and analyze the trajectories of photons in the ATW spacetime. In Sec.~\ref{sec:4}~, we analyze transfer functions when the thin accretion disk is located around the ATW, as well as the observational appearances of the ATW and a BH with the same mass parameter under two emission models. Finally, in Sec.~\ref{sec:5}~, we present our conclusions. Throughout this paper, we employ units in which $G=c=1$.

\section{Effective potential and null geodesic of the asymmetric thin-shell wormhole in Horndeski theory}\label{sec:2}

In this section, inspired by the works~\cite{Wang:2020emr,Guo:2022iiy,Visser:1989kg}, we employ the cut-and-paste method to construct an ATW in Horndeski theory. The ATW consists of two distinct spacetimes, $\mathcal{M}_{1}$ and $\mathcal{M}_{2}$, with different mass parameters. They are glued by a thin shell (i.e.,  ``throat") and the whole  manifold can be labeled as $\mathcal{M}\equiv\mathcal{M}_{1}\cup \mathcal{M}_{2}$.

According to Ref.~\cite{Babichev:2017guv}, we consider a special case of the action (\ref{eq1}), i.e., $G_{2}=\eta X$, $G_{4}=\zeta+\beta\sqrt{-X}$, and $G_{3}=G_{5}=0$. Note that $\eta$ and $\beta$ are dimensionless parameters, but $\zeta=\frac1{2\kappa}$ ($\kappa$ is the Einstein gravitational constant) is just a fixed constant. Hence, the action (\ref{eq1}) can be reduced to
\begin{equation}
	\begin{aligned}\label{eq2}
		S= & \int \mathrm{d}^4 x \sqrt{-g}\left\{\left[\zeta+\beta \sqrt{(\partial \phi)^2 / 2}\right] \mathcal{R}\right.  \left.-\frac{\eta}{2}(\partial \phi)^2-\frac{\beta}{\sqrt{2(\partial \phi)^2}}\left[(\square \phi)^2-\left(\nabla_\mu \nabla_\nu \phi\right)^2\right]\right\}.
	\end{aligned}
\end{equation}
The field equations derived from the action (\ref{eq2}) can give rise to the following BH solution~\cite{Babichev:2017guv}:
\begin{equation}\label{eqds}
	\mathrm{d} s^2=-f(r) \mathrm{d} t^2+\frac{1}{h(r)} \mathrm{d} r^2+r^2\left(\mathrm{d} \theta^2+\sin ^2 \theta \mathrm{d} \phi^2\right)
\end{equation}
with
\begin{equation}\label{eq5}
	f(r)=h(r)=1-\frac{2M}{r}-\frac{\beta^2}{2 \zeta \eta\, r^2},
\end{equation}
where M is a free integration constant representing the mass of the BH. It is worth noting that both parameters $\beta$ and $\eta$ share the same sign, and so the scalar field can be denoted as
\begin{equation}
	\begin{aligned}
		\phi(r)= & \pm 2 \sqrt{\frac{\zeta}{\eta}}\left\{\operatorname{Arctan}\left[\frac{\beta^2+2M\zeta \eta \, r}{\beta \sqrt{2 \zeta \eta \, r(r-2M)-\beta^2}}\right]\!-\operatorname{Arctan}\left(\frac{2M}{\beta} \sqrt{\frac{\zeta \eta}{2}}\right)\right\} \, \text { for } \beta>0   \,\,\& \, \, \eta>0;
	\end{aligned}
\end{equation}
\begin{equation}
	\phi(r)= \pm 2 \sqrt{\frac{\zeta}{-\eta}}\left\{\operatorname{Argth}\left[\frac{\beta^2+2M\zeta \eta \, r}{\beta \sqrt{\beta^2-2 \zeta \eta\, r(r-2M)}}\right]\!+\operatorname{Argth}\left(\frac{2M}{\beta} \sqrt{\frac{-\zeta \eta}{2}}\right)\right\} \, \text { for } \beta<0  \,\,\& \, \, \eta<0.
\end{equation}
Since $\beta^2 /(2 \zeta \eta)$ appears as a whole, for simplicity, we can define the parameter $\gamma=\beta^2 /(2 \zeta \eta)$. In this case, Eq.~(\ref{eq5}) can be re-written as
\begin{equation}\label{eq8}
	f(r)=h(r)=1-\frac{2M}{r}-\frac{\gamma}{r^2}.
\end{equation}
{It can be found that when $\gamma$ is negative, the scalar field $\phi(r)$ exhibits behavior similar to an electric contribution, and $\sqrt{-\gamma}$ resembles an electric charge.} By solving the equation $f(r_h)=0$, the event horizon radius is given by $r_h=M+\sqrt{M^2+\gamma}$. In order to prevent the presence of the naked singularity, it is crucial to ensure that $\gamma \geq -M^2$. When $\gamma = -M^2$, the solution describes an extremal BH.  

Utilizing the cut-and-paste method~\cite{Visser:1989kg}, two distinct Horndeski spacetimes $\mathcal{M}_{1}$ and $\mathcal{M}_{2}$ with different mass parameters can be connected through a thin shell yielding an ATW in Horndeski theory. The metric on the whole manifold $\mathcal{M} \equiv \mathcal{M}_{1} \cup \mathcal{M}_{2}$ (i.e., the ATW spacetime) is given by
\begin{equation}
	d s_i^2=-f_i\left(r_i\right) d t_i^2+\frac{1}{f_i\left(r_i\right)} d r_i^2+r_i^2\left(\mathrm{d} \theta_i^2+\sin ^2 \theta_i \mathrm{d} \phi_i^2\right),
\end{equation}
where
\begin{equation}
	f_i\left(r_i\right)=1-\frac{2 M_i}{r_i}-\frac{\gamma}{r_i^2}, \quad r_i \geq R.
\end{equation}
Here, $i=1,2$ denotes distinct Horndeski spacetimes. The parameter {$R$} represents the position of the throat, which satisfies 
\begin{equation}
	R>\max \left\{M_1+\sqrt{M_1^2+\gamma}, M_2+\sqrt{M_2^2+\gamma}\right\}.
\end{equation}

Considering an incoming photon originating from the spacetime {$\mathcal{M}_{1}$} and passing through the throat, we assume that there is only gravitational interaction between the photon and the throat. This implies that the 4-momentum $p^{a}$ ($a=t,r,\theta,\phi$) of the photon remains constant when passing through the throat. Furthermore, due to the continuity of the metric in the spacetime $\mathcal{M}$, it follows that $g_{\mu\nu}^{\mathcal{M}_{1}}(R) = g_{\mu\nu}^{\mathcal{M}_{2}}(R)$~\cite{Wang:2020emr}. Since the metric~(\ref{eq5}) is static and spherically symmetric, there are two conserved quantities for the photon: $E_i=-p_{t_i}$ (energy) and $L_i=p_{\phi_i}$ (orbital angular momentum). Taking them into the equation of motion of the photon:
\begin{equation}\label{eq3}
	\frac{\left(p_i^{r_i}\right)^2}{f_i\left(r_i\right)}=\frac{p_{t_i}^2}{f_i\left(r_i\right)}-\frac{p_{\phi_i}^2}{r_i^2},
\end{equation}
where $p_i^{a_i}$ represents the 4-momentum of the photon in the spacetime $\mathcal{M}_{i}$, the radial motion of the photon can be expressed as
\begin{equation}\label{eq4}
	p_{i}^{r_{i}}= \pm E_{i} \sqrt{1-\frac{b_{i}^2}{r_{i}^2} f_{i}\left(r_{i}\right)}.
\end{equation}
The symbol $\pm$ corresponds to the directions of the outgoing and incoming photons, respectively. The parameter $b_i = L_i / E_i$ denotes the impact parameter of the photon in the spacetime $\mathcal{M}_{i}$. With Eq.~(\ref{eq4}), the effective potential $V_{i}(r_{i})$ of the ATW can be given by
\begin{equation}
	V_{i}(r_{i})=\frac{f_{i}\left(r_{i}\right)}{r_{i}^2}=\frac{1}{r_{i}^2}\left(1-\frac{2 M_i}{r_i}-\frac{\gamma}{r_i^2}\right).
\end{equation}
For the unstable circular orbit of the photon, the effective potential satisfies the following conditions:
\begin{equation}\label{eq10}
	V_{i}\left(r_{ph_{i}}\right)=\frac{1}{b_{c_{i}}^2}, \quad V_{i}^{\prime}(r_{ph_{i}})=0,
\end{equation}
where $r_{ph_{i}}$ is the radius of the photon sphere and $b_{c_{i}}$ is the critical impact parameter. Note that these two parameters only depend on $M_i$ and $\gamma$. 

In this study, we focus on how to distinguish between ATWs and BHs in Horndeski theory when the event horizon is obscured by a photon sphere. We can suppose that the observer is situated in the spacetime $\mathcal{M}_{1}$ with $M_1=1$ and the mass parameter of the spacetime $\mathcal{M}_{2}$ can be set to $M_2=k$. According to Refs.~\cite{Wang:2020emr,Peng:2021osd}, $k$ and $R$ should satisfy 
\begin{equation}\label{eq18}
	1<k<\frac{R}{2} \leq \frac{r_{ph_{1}}}{2}.
\end{equation}
In fact, for any $1<k<\frac{R}{2}$, our subsequent discussion and results are similar. Therefore, we follow the practice adopted in Refs.~\cite{Peng:2021osd,Guo:2022iiy} and let $M_2=k=1.2$ for simplicity. The impact parameters $b_{1}$ and $b_{2}$ in the spacetimes $\mathcal{M}_{1}$ and $\mathcal{M}_{2}$ can be connected through the following equation~\cite{Wang:2020emr}:   
\begin{equation}\label{eqz}
	\frac{b_1}{b_2}=\sqrt{\frac{f_2(R)}{f_1(R)}}=\sqrt{\frac{1-\frac{2 M_2}{R}-\frac{\gamma}{R^2}}{1-\frac{2 M_1}{R}-\frac{\gamma}{R^2}}} \equiv Z.
\end{equation}
For the given $M_1=1$ and $M_2=1.2$, to study the influence of the parameter $\gamma$ on the values of $r_{ph_{i}}$ and $b_{c_{i}}$, we use Eq.~(\ref{eq10}) to calculate $r_{ph_i}$ and $b_{c_{i}}$ for various positive and negative values of $\gamma$, as shown in Table~\ref{TAB}~. It is worth noting that $\gamma=0$ and $\gamma<0$ correspond to the Schwarzschild case and the Reissner-Nordstr$\ddot{\text{o}}$m case, respectively. The results demonstrate a direct correlation between an increase in the parameter $\gamma$ and increases in both $r_{ph_i}$ and $b_{c_i}$.
\begin{table}[h!]
	\centering
	\begin{tabular}{|l|r|c|c|c|c|c|c|c}
		\hline
		$~~\gamma$ & $-0.2~~~$ & $-0.1$ & $0$ & $0.1$ & $0.2$ \\
		\hline
		$~~b_{c_{1}}$ & $5.01561$ & $5.10779$ & $5.19615$ & $5.28114$ & $5.36311$ \\
		\hline
		$~~r_{ph_{1}}$ & $2.86015$ & $2.93178$ & $3$ & $3.06525$ & $3.12788$ \\
		\hline
		$~~b_{c_{2}}$ & $6.08691$ & $6.16221$ & $6.23538$ & $6.30659$ & $6.37602$\\
		\hline
		$~~r_{ph_{2}}$ & $3.48523$ & $3.54356$ & $3.6$ & $3.65472$ & $3.70788$ \\
		\hline
	\end{tabular}
	\caption{The radius $r_{ph_{i}}$ of the photon sphere and the critical impact parameter $b_{c_{i}}$. We consider five specific values for the parameter $\gamma$: $-0.2,\,-0.1,\,0,\,0.1,\,0.2$. For $r_{ph_{1}}$ and $b_{c_{1}}$, the mass parameter is $M_1=1$, while for $r_{ph_{2}}$ and $b_{c_{2}}$, the mass parameter is $M_2=1.2$.}\label{TAB}
\end{table}

In Fig.~\ref{potential1}~, we plot the effective potentials of ATWs and BHs for different values of $\gamma$. According to Eq.~(\ref{eqz}), we rescale the effective potential $V_{2}(r_{2})$ in the spacetime $\mathcal{M}_2$ by a factor of $Z^2$, as shown in Fig.~\ref{fig:eff2}~. It is found that for a given $\gamma$, the trajectories of photons can be classified into three scenarios. For example, when $\gamma=-0.2$ (see the blue and green lines), if $b_1>b_{c_1}=5.01561$, the photons in spacetime $\mathcal{M}_1$ will approach turning points in the spacetime $\mathcal{M}_1$ from infinity and then return to infinity in spacetime $\mathcal{M}_1$, i.e., the photons remain within the spacetime $\mathcal{M}_1$ throughout their trajectories. If $Zb_{c_2}=3.89321<b_1<b_{c_1}^c$, the photons in the spacetime $\mathcal{M}_1$ will reach turning points in the spacetime $\mathcal{M}_2$ and then return to the spacetime $\mathcal{M}_1$. If $b_1<Zb_{c_2}$, the photons in spacetime $\mathcal{M}_1$ will fall into the spacetime $\mathcal{M}_2$ and then go to infinity in the spacetime $\mathcal{M}_2$.

For the effective potentials of BHs [see Fig.~\ref{fig:eff1}~], one can find that a light ray's trajectory can also be classified into three distinct cases~\cite{Gralla:2019xty} [see the illustration positioned in the lower right corner in Fig.~\ref{fig:eff1}~]. When the impact parameter of a photon is less than the critical impact parameter $b_{c}$ (for example see Region 3 for $\gamma = -0.2$ and $b_{c}$ = $5.01561$), it will fall into the BH. When the impact parameter of a photon is equal to the critical value $b_{\rm c}$ (for example see Region 2 for $\gamma = -0.2$ and $b_{c}$ = $5.01561$), it can undergo periodic circular motion around the BH. When the impact parameter of a photon is larger than $b_{c}$ (for example see Region 1 for $\gamma = -0.2$ and $b_{c}$ = $5.01561$), the gravitational force will cause the deflection of the light ray's trajectory. Moreover, Fig.~\ref{fig:eff1} illustrates that the peak of the effective potential increases with the parameter $\gamma$.

\begin{figure}[]
	\centering
	\subfigure[\label{fig:eff2}]{
		\begin{minipage}[t]{0.5\linewidth}
			\centering
			\includegraphics[width=6cm,height=6cm]{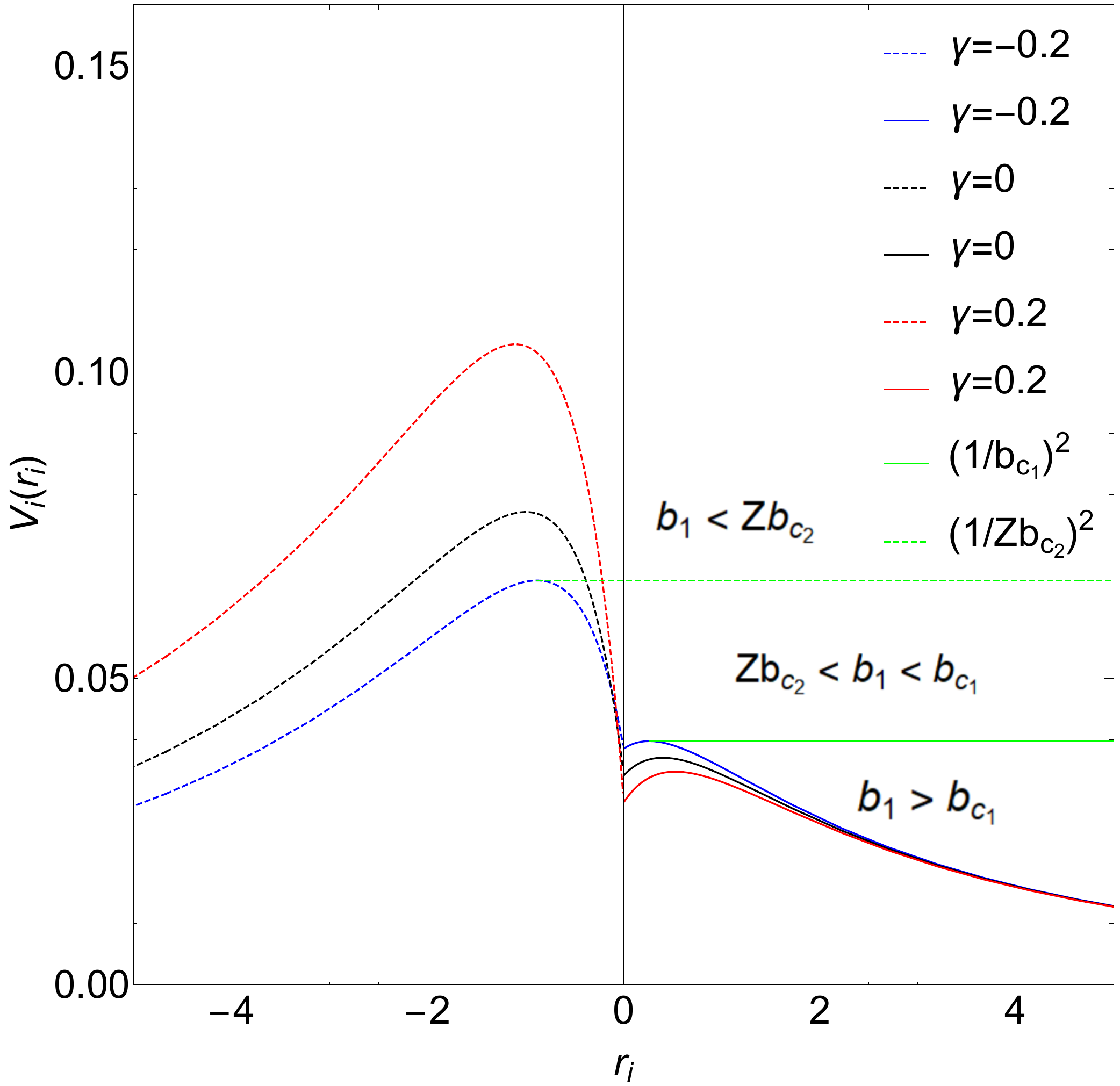}
		\end{minipage}%
	}%
	\subfigure[\label{fig:eff1}]{
		\begin{minipage}[t]{0.5\linewidth}
			\centering
			\includegraphics[width=6cm,height=6cm]{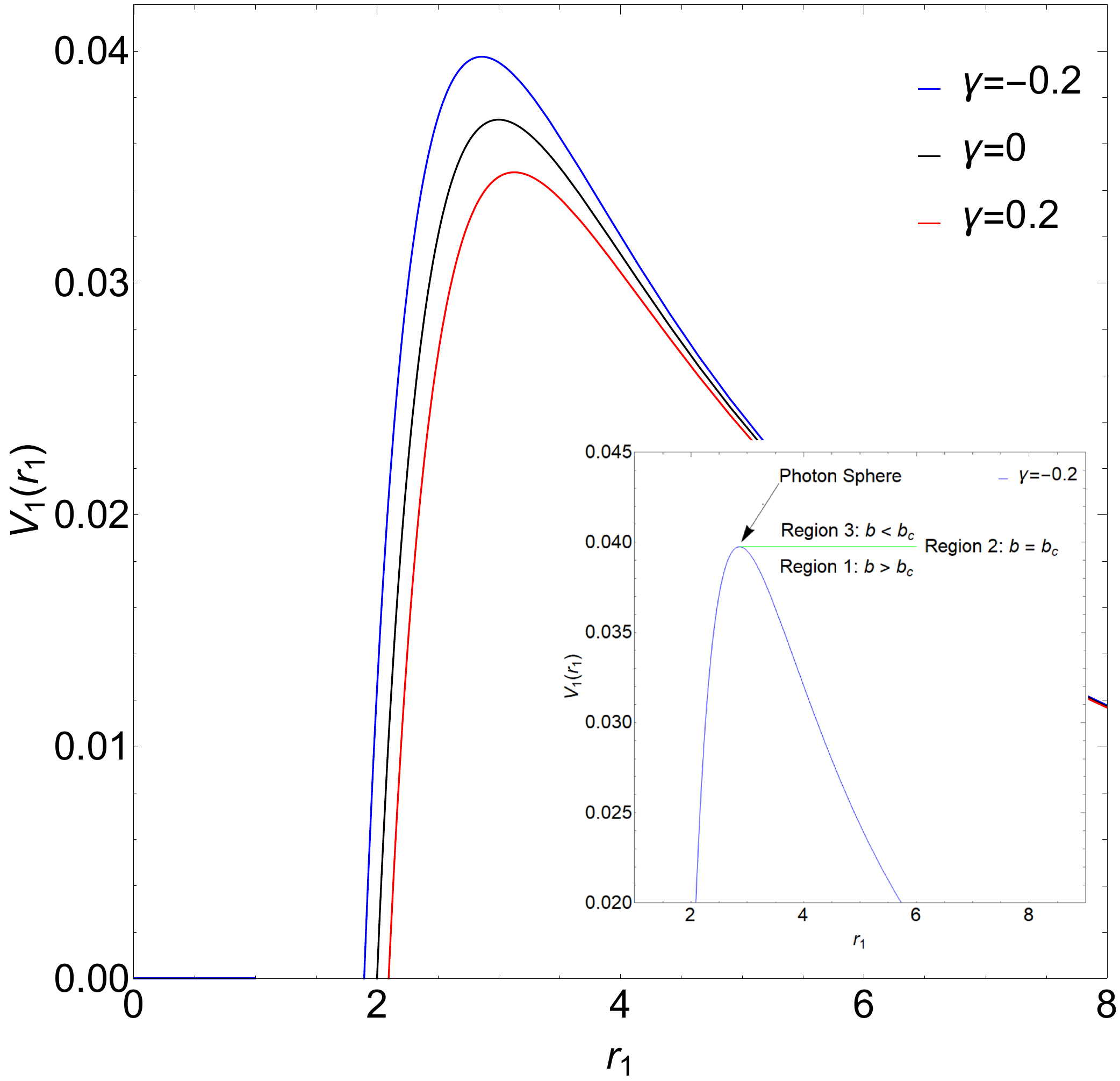}
		\end{minipage}%
	}
	\caption{The effective potentials of ATWs and BHs. We consider three specific values for the parameter $\gamma$: $-0.2,\,0,\,0.2$. The left panel is for ATWs, where the solid and dashed lines represent the effective potentials of the spacetime $\mathcal{M}_1$ and the spacetime $\mathcal{M}_2$, respectively. {When $\gamma=-0.2$, we have $b_{c_1}=5.01561$ and $Zb_{c_2}=3.89321$ (see the green lines).} The right panel is for BHs. We set $M_1=1,\, M_2=1.2$, and $R=2.6$.} 
	\label{potential1}
\end{figure}

To get more information about the observational appearance of the ATW in Horndeski theory, it is necessary to study the trajectories and deflection angles of photons as they propagate in the ATW spacetime. With Eq. (\ref{eq3}), the trajectory of a photon in the ATW spacetime can be formulated as
\begin{equation}
	\frac{1}{b_{i}^2}-\frac{f_{i}\left(r_{i}\right)}{r_{i}^2}=\frac{1}{r_{i}^4}\left(\frac{\mathrm{d} r_{i}}{\mathrm{d} \phi_{i}}\right)^2.
\end{equation}
By implementing the coordinate transformation $x \equiv 1/r$, one can obtain
\begin{equation}\label{geo}
	G_i\left(x_i\right)=\left(\frac{d x_i}{d \phi}\right)^2=\frac{1}{b_i^2}-x_i^2\left(1-2 M_i x_i-x_i^2 \gamma\right).
\end{equation}
When $b_1>b_{c_1}$, the photon remains in the spacetime $\mathcal{M}_1$. The turning point relies on the smallest positive root of $G_1(x_1)=0$, which is denoted as $x_1^{min}$. Referring to Eq.~\eqref{geo}, the deflection angle (the total change in the azimuth angle) of the photon in the spacetime $\mathcal{M}_1$ is
\begin{equation}\label{eq20}
	\phi_1\left(b_1\right)=2 \int_0^{x_1^{m i n}} \frac{d x_1}{\sqrt{G_1\left(x_1\right)}}, \quad b_1>b_{c_1} .
\end{equation}
When $Zb_{c_2}<b_1<b_{c_1}$, the photon could reach the turning point in the spacetime $\mathcal{M}_2$ and return to the spacetime $\mathcal{M}_1$. The deflection angle of the photon in the spacetime $\mathcal{M}_1$ is 
\begin{equation}\label{phi2}
	\phi_1^*\left(b_1\right)=\int_0^{1/R} \frac{d x_1}{\sqrt{G_1\left(x_1\right)}}, \quad b_1<b_{c_1}.
\end{equation}
The turning point in the spacetime $\mathcal{M}_2$ is determined by the largest positive root of $G_2(x_2)=0$, which is denoted as $x_2^{max}$. In this case, the impact parameter $b_2$ of the photon in the spacetime $\mathcal{M}_2$ is given by Eq.~(\ref{eqz}). Therefore, the deflection angle of the photon in the spacetime $\mathcal{M}_2$ is 
\begin{equation}\label{phi3}
	\phi_2\left(b_2\right)=2 \int_{x_2^{max}}^{1/R} \frac{d x_2}{\sqrt{G_2\left(x_2\right)}}, \quad b_2>b_{c_2} .
\end{equation}
When $b_1<Zb_{c_2}$, since the photon  will drop into the spacetime $\mathcal{M}_2$ and then move to infinity in the spacetime $\mathcal{M}_2$, we do not need to consider the deflection angle of the photon.  
 
Using Eqs.~(\ref{eq20})-(\ref{phi3}), we can plot the trajectories of photons in the ATW spacetime for different values of the parameter $\gamma$ and the impact parameter $b_1$. The results are illustrated in Fig.~\ref{trajectory}~. Here, we take the case of $\gamma=-0.2$ [see Figs.~\ref{trajectory1}~, \ref{trajectory2}~, and \ref{trajectory3}~] as an example to the trajectories of photons when $Zb_{c_2}<b_1<b_{c_1}$. For a light ray originates from infinity in the spacetime $\mathcal{M}_{1}$, it can be seen that when the impact parameter $b_1$ decreases, the light trajectory thin the spacetime $\mathcal{M}_{2}$ will be longer. From the left column and the right column in Fig. \ref{trajectory}~, one can find that $b_{c_1}$ increases with the parameter $\gamma$, while $Zb_{c_2}$ decreases with it.

\begin{figure}[h!]
	\centering
	\subfigure[$\gamma=-0.2,~b_{c_1}=5.01561$\label{trajectory1}]{
		\begin{minipage}[t]{0.33\linewidth}
			\centering
			\includegraphics[width=4cm,height=4cm]{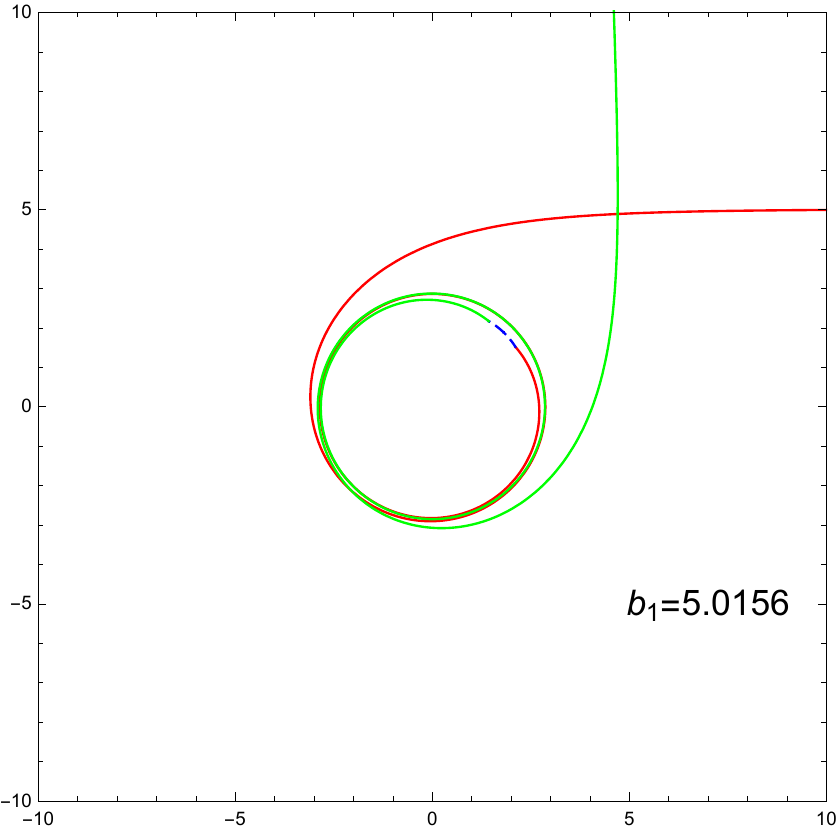}
		\end{minipage}%
	}%
	\subfigure[$\gamma=-0.2
	$\label{trajectory2}]{
		\begin{minipage}[t]{0.33\linewidth}
			\centering
			\includegraphics[width=4cm,height=4cm]{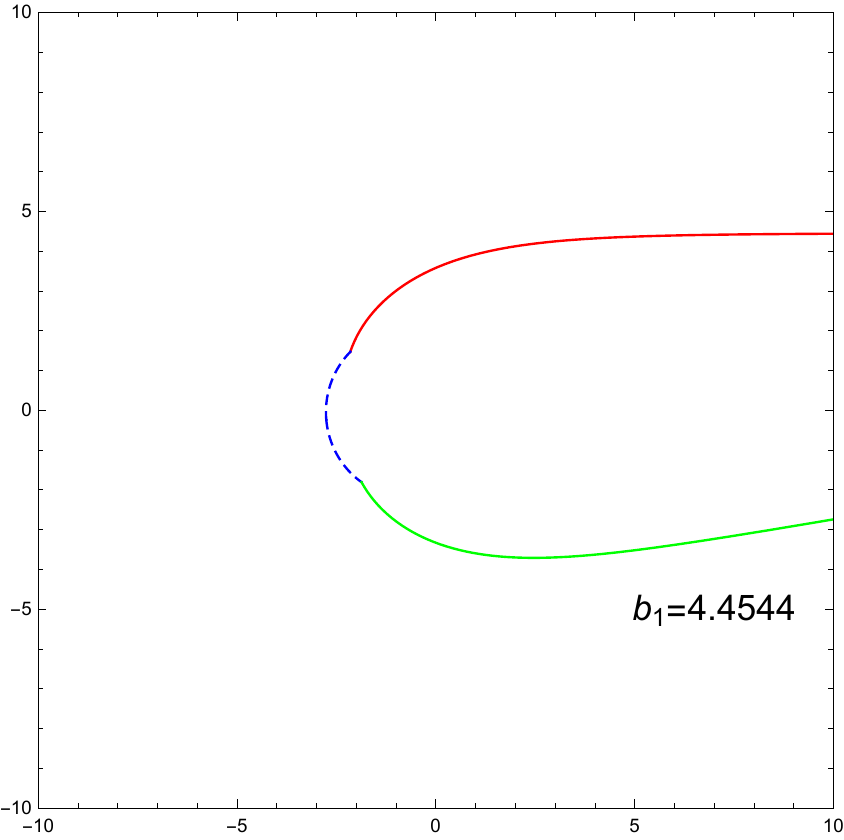}
		\end{minipage}%
	}%
	\subfigure[$\gamma=-0.2,~Zb_{c_2}=3.89320$\label{trajectory3}]{
		\begin{minipage}[t]{0.33\linewidth}
			\centering
			\includegraphics[width=4cm,height=4cm]{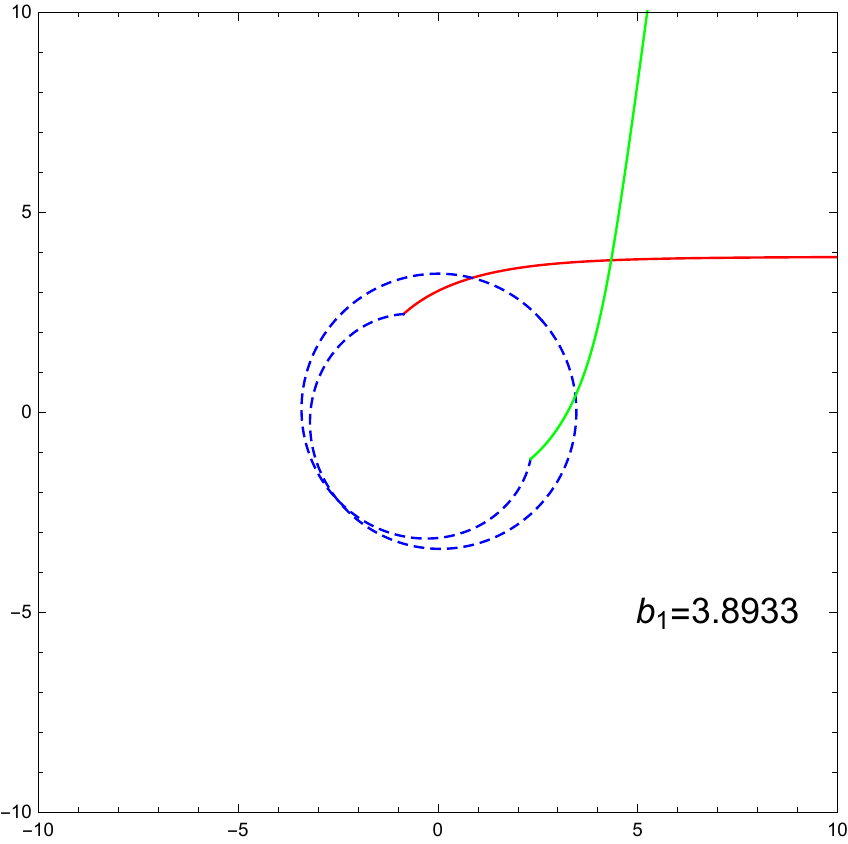}
		\end{minipage}
	}%
	\quad
	\subfigure[$\gamma=0,~b_{c_1}=5.19615$\label{trajectory4}]{
		\begin{minipage}[t]{0.32\linewidth}
			\centering
			\includegraphics[width=4cm,height=4cm]{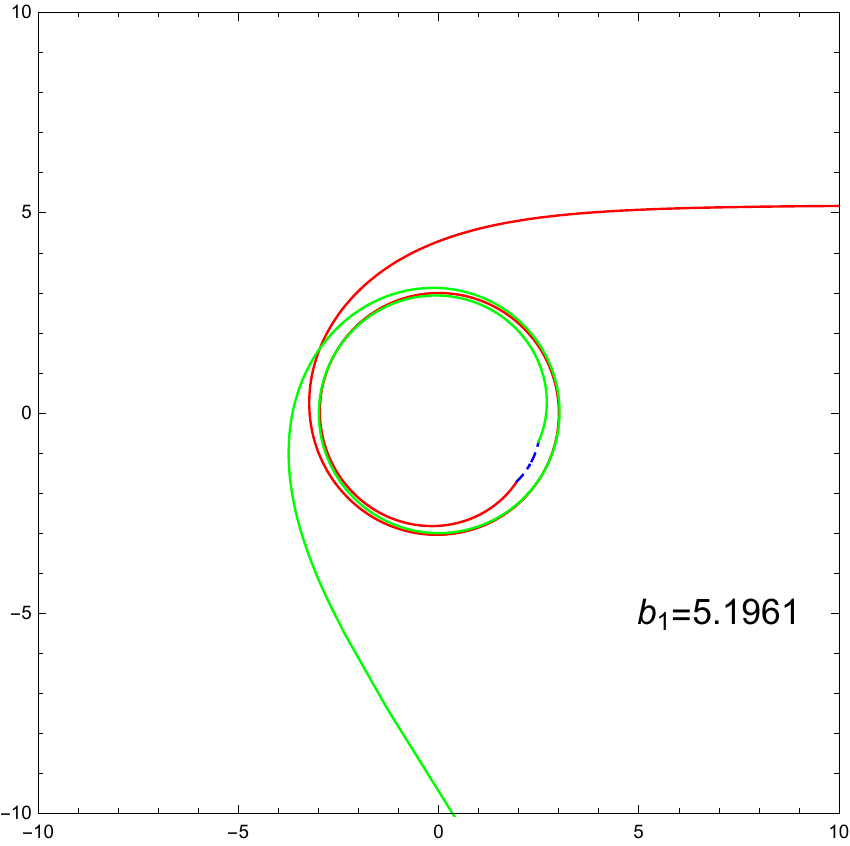}
		\end{minipage}
	}%
	\subfigure[$\gamma=0$\label{trajectory5}]{
		\begin{minipage}[t]{0.32\linewidth}
			\centering
			\includegraphics[width=4cm,height=4cm]{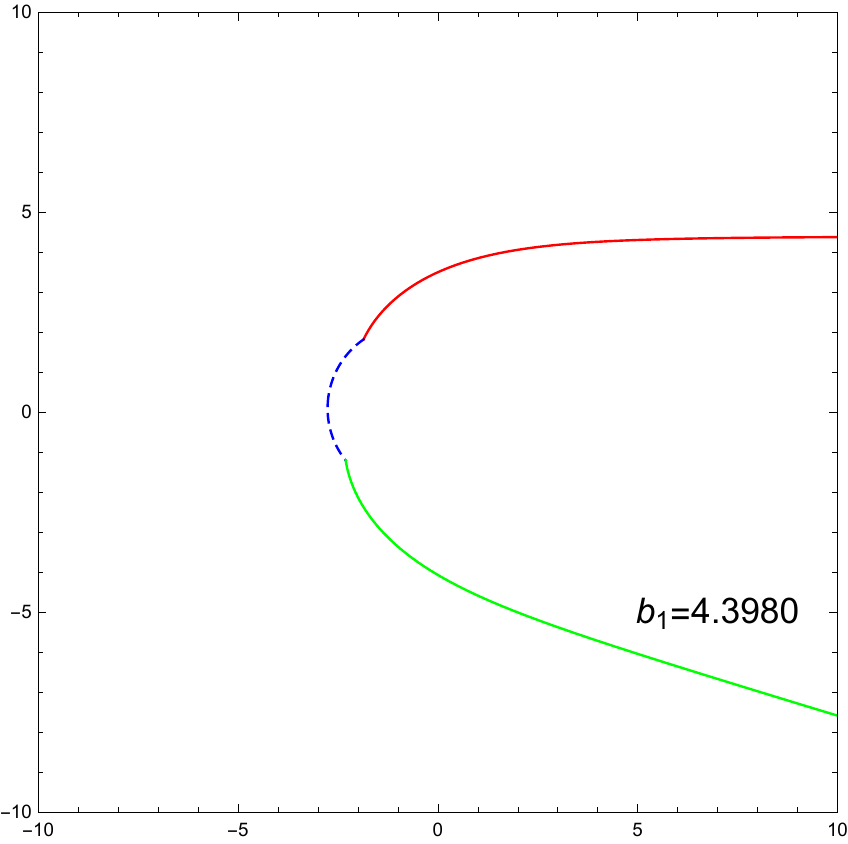}
		\end{minipage}
	}%
	\subfigure[$\gamma=0,~Zb_{c_2}=3.6$\label{trajectory6}]{
		\begin{minipage}[t]{0.32\linewidth}
			\centering
			\includegraphics[width=4cm,height=4cm]{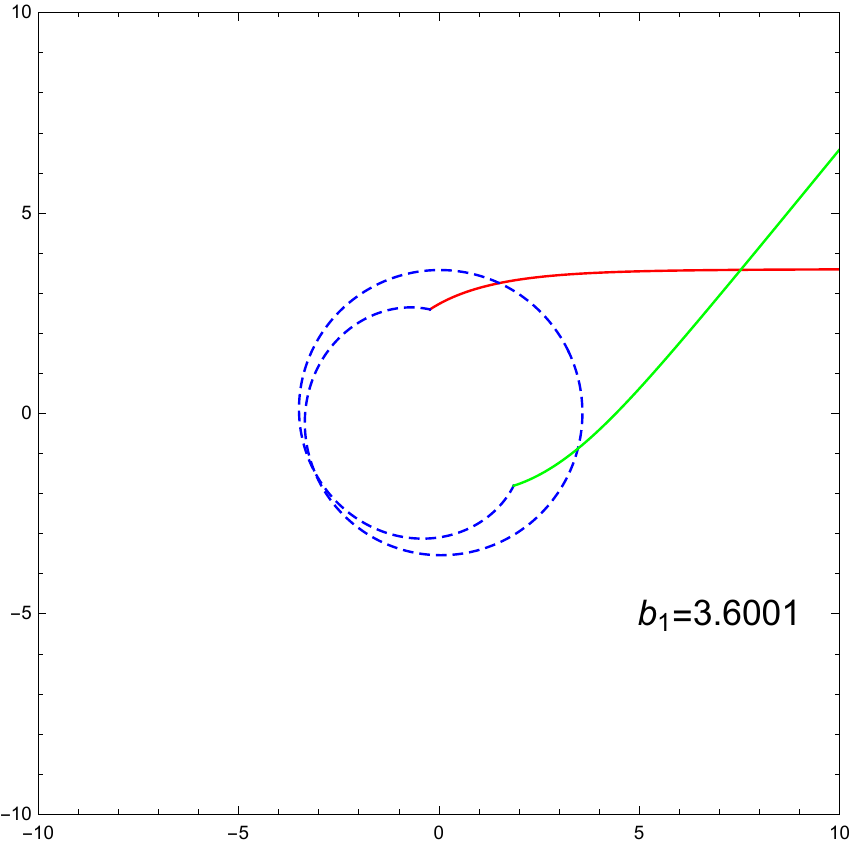}
		\end{minipage}
	}%
	\quad
	\subfigure[$\gamma=0.2,~b_{c_1}=5.36311$\label{trajectory7}]{
		\begin{minipage}[t]{0.32\linewidth}
			\centering
			\includegraphics[width=4cm,height=4cm]{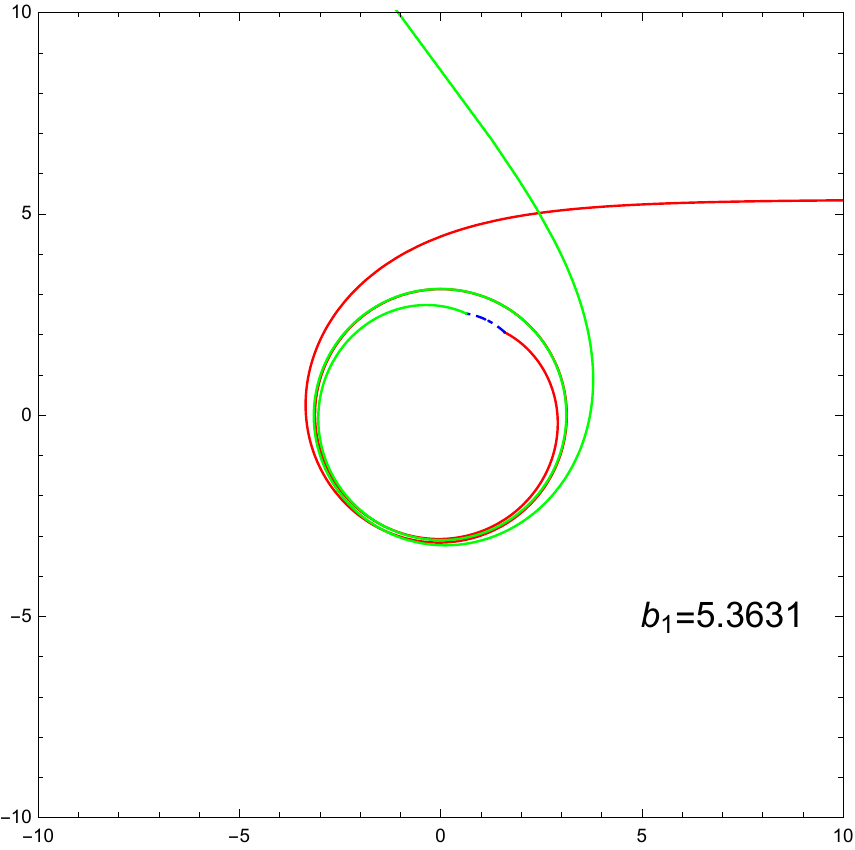}
		\end{minipage}
	}%
	\subfigure[$\gamma=0.2$\label{trajectory8}]{
		\begin{minipage}[t]{0.32\linewidth}
			\centering
			\includegraphics[width=4cm,height=4cm]{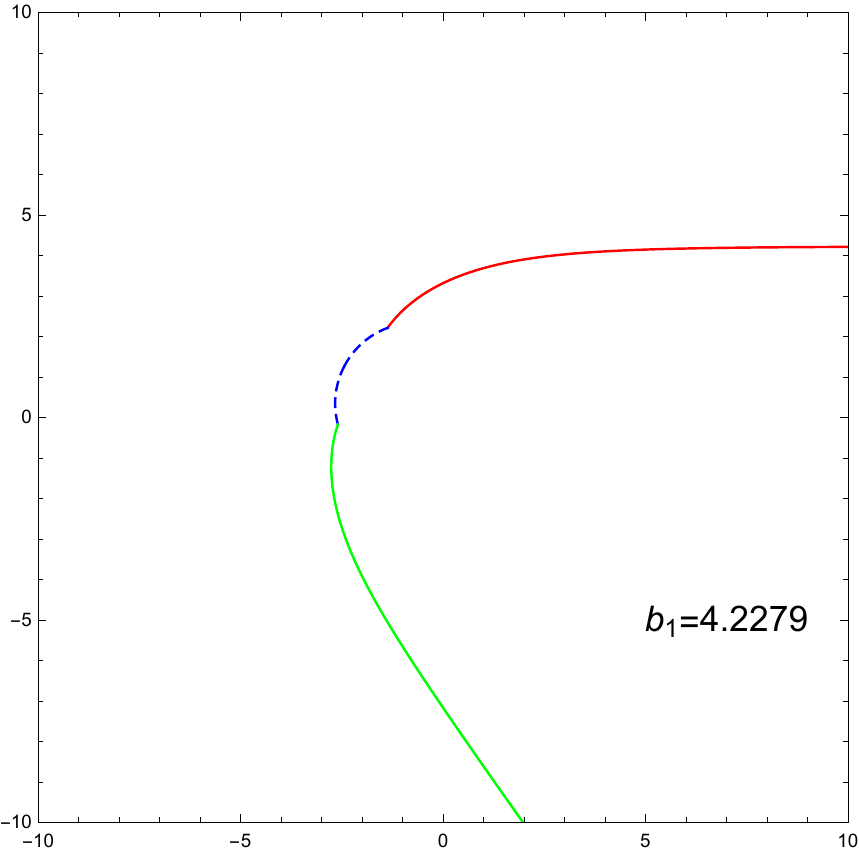}
		\end{minipage}
	}%
	\subfigure[$\gamma=0.2,~Zb_{c_2}=3.09283$\label{trajectory9}]{
		\begin{minipage}[t]{0.32\linewidth}
			\centering
			\includegraphics[width=4cm,height=4cm]{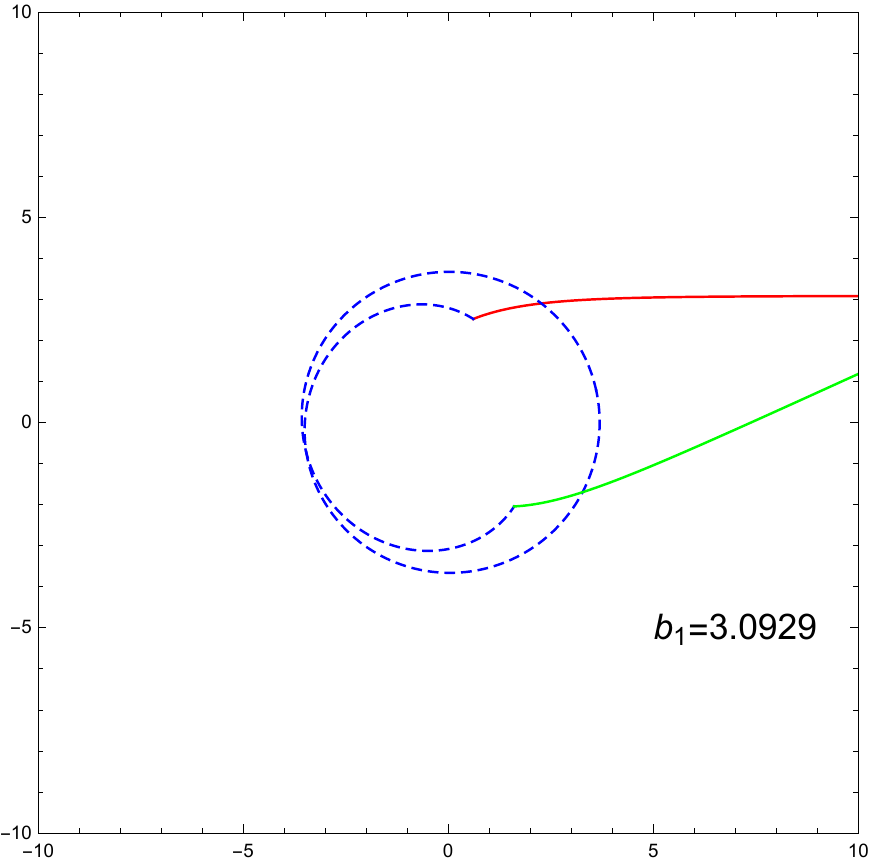}
		\end{minipage}
	}%
	
	\caption{The trajectories of photons in the polar coordinate $(r_{1}, \phi)$ with the impact parameter $Z b_{c_2}<$ $b_1<b_{c_1}$. The upper, middle, and lower panels correspond to $\gamma=-0.2$, $\gamma=0$, and $\gamma=0.2$, respectively. The incoming and outgoing light trajectories in the spacetime $\mathcal{M}_{1}$ are represented by the red solid lines and the green solid lines, respectively, while the light trajectories in the spacetime $\mathcal{M}_{2}$ indicated by the blue dashed lines. We set $M_1=1, M_2=1.2$, and $R=2.6$.}
	\label{trajectory}
\end{figure}

\section{Observational appearance of the asymmetric thin-shell wormhole}
\label{sec:4}

Since the reflection mechanism of ATWs differentiates them from that of BHs, the observable appearance of ATWs is different from that of BHs. In this section, we consider the emission originating from an optically and geometrically thin disk encircling an ATW in Horndeski theory. By comparing the observational appearances of the ATW and a BH with the same mass parameter, we can find that there exist distinct properties in the observational appearance of the ATW.

\subsection{\textbf{Classification of light trajectories}}
\label{sec:4-1}

 In Ref.~\cite{Gralla:2019xty}, Gralla $et~al.$ investigated the observational appearance of a BH surrounded by a thin accretion disk by considering the orbit number ($n = \phi / 2 \pi$) and the number of intersections between light rays and the accretion disk. Based on the orbit number, the light trajectories can be classified into three categories:
\begin{itemize}
	\item \emph{Direct emission:}~~ $n < 3/4$. A light ray intersects with the equatorial plane only once;
	\item \emph{Lensing ring:}~~ $3/4 < n < 5/4$. A Light rays crosses the equatorial plane twice;
	\item \emph{Photon ring:}~~ $n > 5/4$. A light ray crosses the equatorial plane at least three times.
\end{itemize}

In the context of an ATW with a thin accretion disk, we consider an observer located at the North Pole in the spacetime $\mathcal{M}_{1}$. When a light ray originates from the spacetime $\mathcal{M}_{1}$ and subsequently drops into the spacetime $\mathcal{M}_{2}$ through the throat, we can redefine the orbit number as follows\cite{Peng:2021osd}:
\begin{eqnarray}
	\label{4-1}
	&&n_{1}(b_{1}) = \frac{\phi_{1}(b_{1})}{2\pi},\\
	&&n_{2}(b_{2}) = \frac{\phi_{1}^{*}(b_{1}) + \phi_{2}(b_{1}/Z)}{2\pi},\\
	&&n_{3}(b_{1}) = \frac{2 \phi_{1}^{*}(b_{1}) + \phi_{2}(b_{1}/Z)}{2\pi}.
\end{eqnarray}
Here, $n_{2}$ and $n_{3}$ are extra orbit functions for the ATW, which correspond to extra photon rings. 

The correlation between the orbit number and the impact parameter $b_1$ of the light ray is shown in Fig.~\ref{transfer1}~. Here $n_{1}$ is the similar to the orbit number of a BH because the light ray remains in the spacetime $\mathcal{M}_{1}$ [see Fig. \ref{fig:n1}~]. For the light ray dropping into the spacetime $\mathcal{M}_{2}$ and the returning to the spacetime $\mathcal{M}_{1}$, if $n_{2}<3/4$ [the solid lines in Fig. \ref{fig:n2n3}~] and $n_{3}>3/4$ [the dashed lines in Fig. \ref{fig:n2n3}~], the light ray finally intersects with the accretion disk on the back side of the accretion disk, while if $n_{2}<5/4$ and $n_{3}>5/4$, the light ray finally intersects with the accretion disk on the front side of the accretion disk. Moreover, Fig.~\ref{fig:n2n3} indicates that the range of the impact parameter $b_1$ increases with the value of $\gamma$, which aligns with the results shown in Table~\ref{TAB} and Fig.~\ref{trajectory}~. This implies that as $\gamma$ increases, these extra photon rings will expand towards the outer region of the ATW shadow.

\begin{figure}[!b]
	\centering
	\subfigure[Orbit number $n_1$\label{fig:n1}]{
		\begin{minipage}[t]{0.5\linewidth}
			\centering
			\includegraphics[width=6cm,height=6cm]{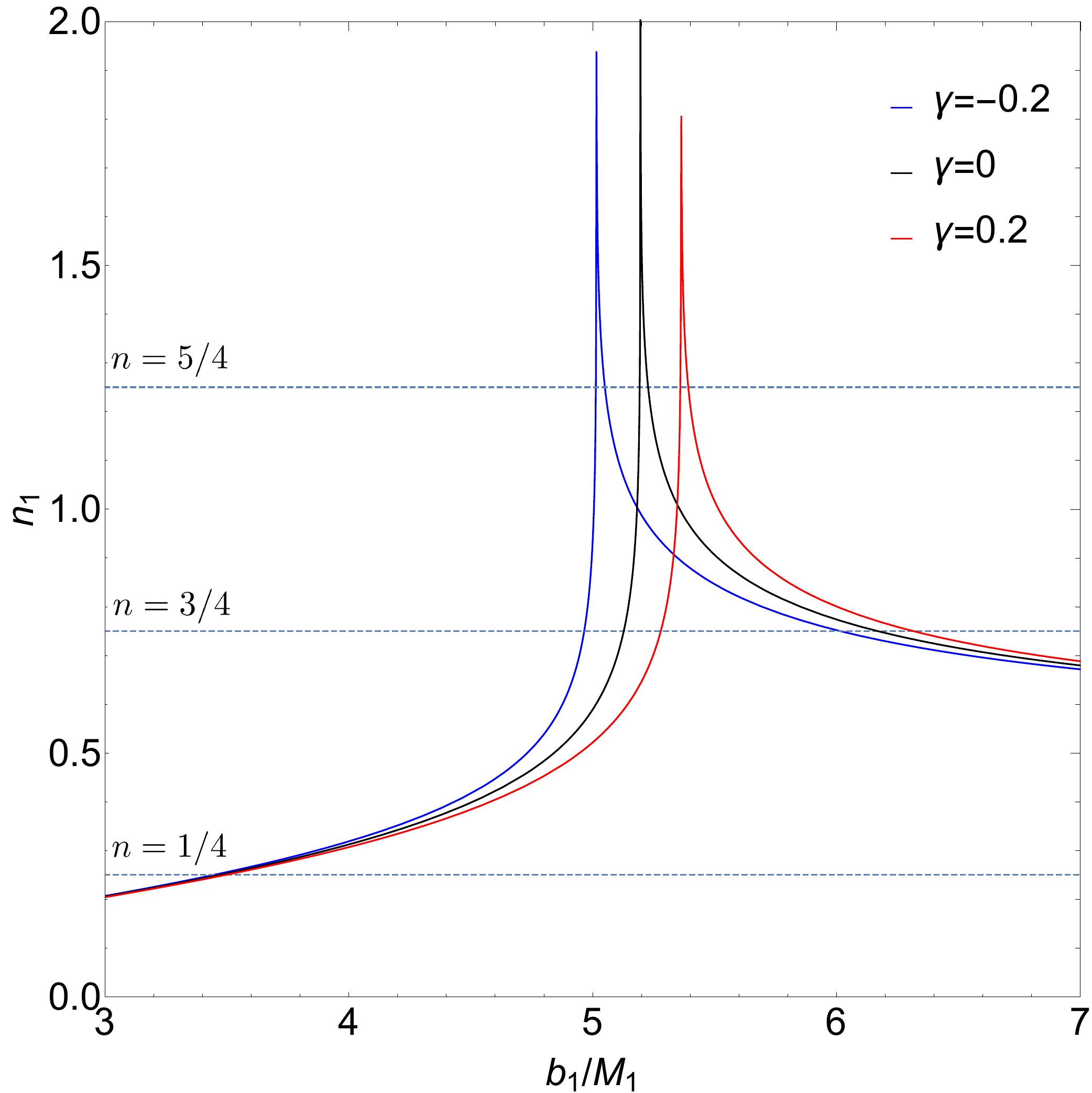}
		\end{minipage}%
	}%
\subfigure[Orbit number $n_2$,~$n_3$\label{fig:n2n3}]{
		\begin{minipage}[t]{0.5\linewidth}
			\centering
			\includegraphics[width=6cm,height=6cm]{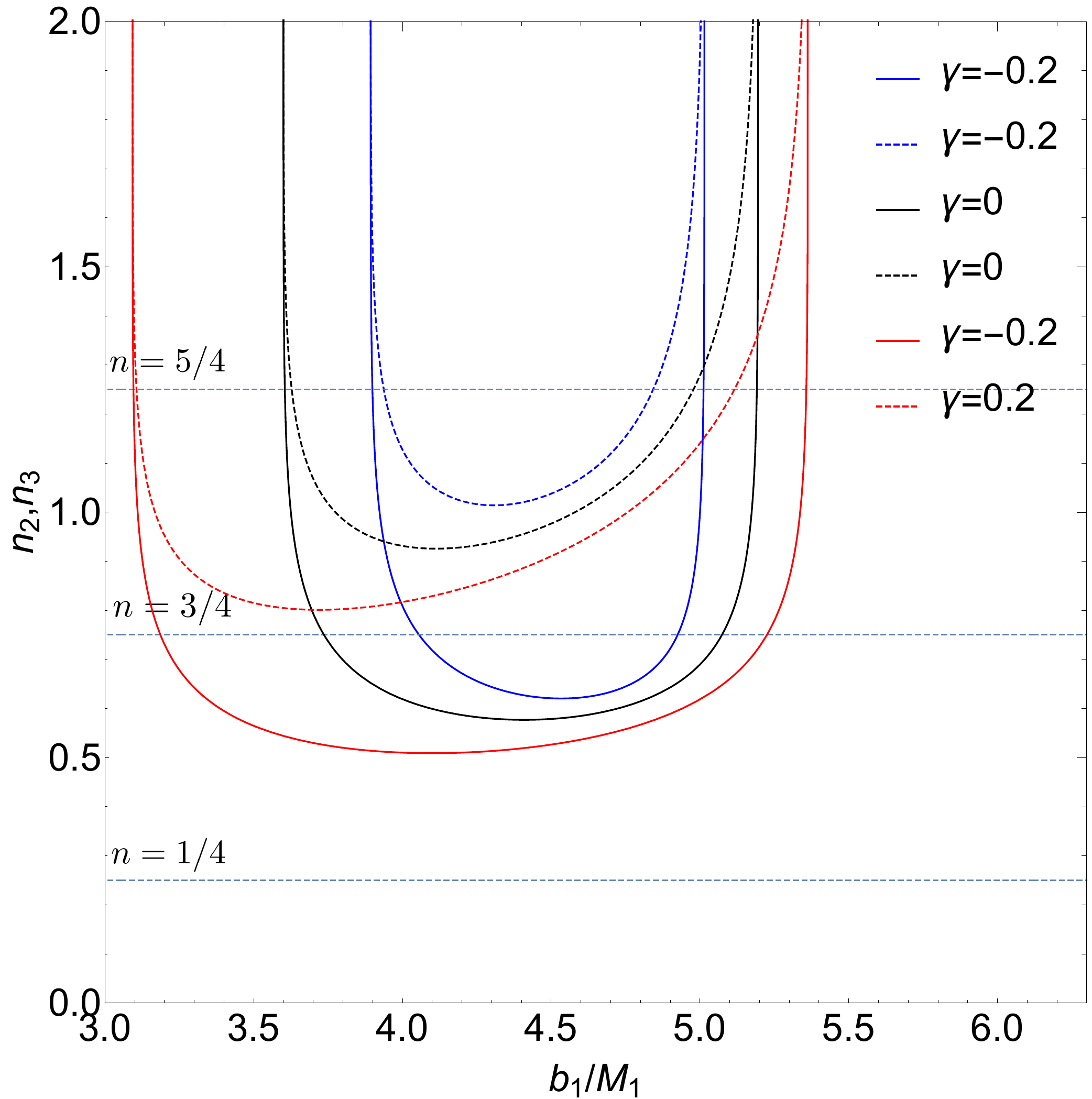}
		\end{minipage}%
	}
	\caption{The orbit number of the light rays around ATWs. The blue, black, and red lines correspond to $\gamma=-0.2$, $\gamma=0$, and $\gamma=0.2$, respectively.	In the right panel, the solid lines represent $n_2$, while the dashed lines represent $n_3$.	We set $M_1=1, M_2=1.2$, and $R=2.6$.}
	\label{transfer1}
\end{figure}

\subsection{\textbf{Observed intensity and transfer functions}}
\label{sec:4-2}
In this part, we consider that the emission originates from an optically and geometrically thin accretion disk around an ATW located in the spacetime $\mathcal{M}_{1}$. We assume that the static observer in the spacetime $\mathcal{M}_{1}$ is at the North Pole, and the disk is situated in the equatorial plane. In the rest frame of the static observer, the disk emits light rays isotropically. Due to the spherical symmetry of the spacetime, the emitted specific intensity is only related to the radial coordinate, represented as $I^{\rm em}_\nu(r)$, where $\nu$ is the emission frequency in the static frame. If the static observer at infinity receives a specific intensity $I^{\rm obs}_{\nu'}$ with a redshifted frequency $\nu'=\sqrt{f}\nu$, then we have
\begin{equation}
	\frac{I_{\nu^{\prime}}^{\mathrm{obs}}}{\nu^{\prime 3}}=\frac{I_\nu^{\mathrm{em}}}{\nu^3}.
\end{equation}
Then the observed specific intensity can be rewritten as
\begin{equation}\label{iobsp}
	I_{\nu^{\prime}}^{\mathrm{obs}}=f^{3 / 2}(r) I_\nu^{\mathrm{em}}(r).
\end{equation} 
By integrating Eq.~(\ref{iobsp}) over all frequencies, one can obtain the total observed intensity for each intersection:
\begin{equation}
	I^{\mathrm{obs}}=\int I_{\nu^{\prime}}^{\mathrm{obs}} d \nu^{\prime}=\int f^2 I_\nu^{\mathrm{em}} d \nu=f^2(r) I^{\mathrm{em}}(r),
\end{equation}
in which $I^{\mathrm{em}}(r)=\int I_\nu^{\mathrm{em}} d \nu$ represents the total emitted intensity of the accretion disk. Therefore, the total observed intensity for all intersections can be calculated as follows:
\begin{equation}\label{iobs}
	I^{\text {obs }}(b)=\left.\sum_n I^{\mathrm{em}}(r) f^2(r)\right|_{r=r_n(b_1)},
\end{equation}
where the transfer function $r_{n}(b_1)$ determines the radial location of the $n$-th intersection between the accretion disk and the light ray with the impact parameter $b_1$. According to Ref.~\cite{Gralla:2019xty}, the demagnification factor can be defined as the derivative of the transfer function, i.e., ${\rm d}r/{\rm d}b_1$. The first transfer function ($n=1$) provides the “direct emission” of the accretion disk, while the second transfer function ($n=2$) and the third transfer function ($n=3$) correspond to the ``lensing ring" and the ``photon ring", respectively.   

The transfer functions with respective to the impact parameter $b_1$ are plotted in Fig.~\ref{transfer2}~. We consider three specific values for the parameter $\gamma$: $-0.2,\,0,\,0.2$. It can be found that the ``direct image" (the redshift of the source profile) of the accretion disk is given by the first transfer function ($n=1$, the black lines), which has a small demagnification factor. The second transfer function ($n=2$, the blue lines) produces a ``lensing ring" with a large demagnification factor, which represents a reduced image of the back side of the accretion disk. The third transfer function ($n=3$, the red lines) exhibits the largest demagnification factor and generates a highly reduced image known as the ``photon ring" on the front side of the accretion disk. Compared to the BH case, the blue dashed line ($n=2$) represents the new second transfer function of the ATW, which corresponds to the ``lensing band"~\cite{Peng:2021osd}. The red dashed lines near $Zb_{c_2}$ and $b_{c_1}$ are the new third transfer functions that correspond to the ``photon ring group"~\cite{Guo:2022iiy}. Furthermore, comparing the demagnification factors in the three cases of $\gamma$ [see the dashed blue lines in Fig.~\ref{transfer2}~], it is evident that an increase in the parameter $\gamma$ leads to a higher demagnification factor for the new second transfer function.

\begin{figure}[!ht]
	\centering
	\subfigure[$\gamma=-0.2$\label{transferI}]{
		\begin{minipage}[t]{0.33\linewidth}
			\centering
			\includegraphics[width=4cm,height=4cm]{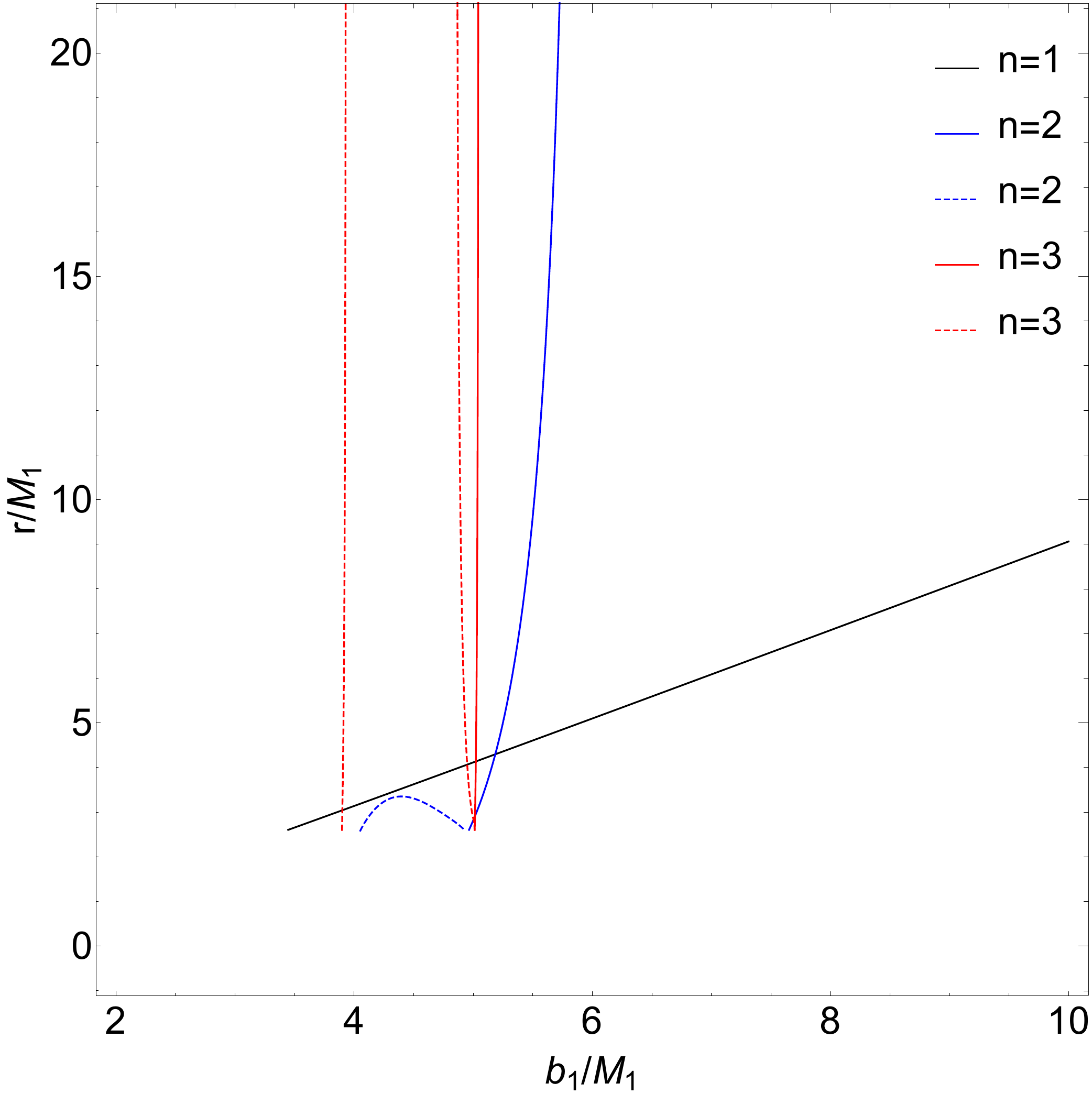}
		\end{minipage}%
	}%
	\subfigure[$\gamma=0
	$\label{transferII}]{
		\begin{minipage}[t]{0.33\linewidth}
			\centering
			\includegraphics[width=4cm,height=4cm]{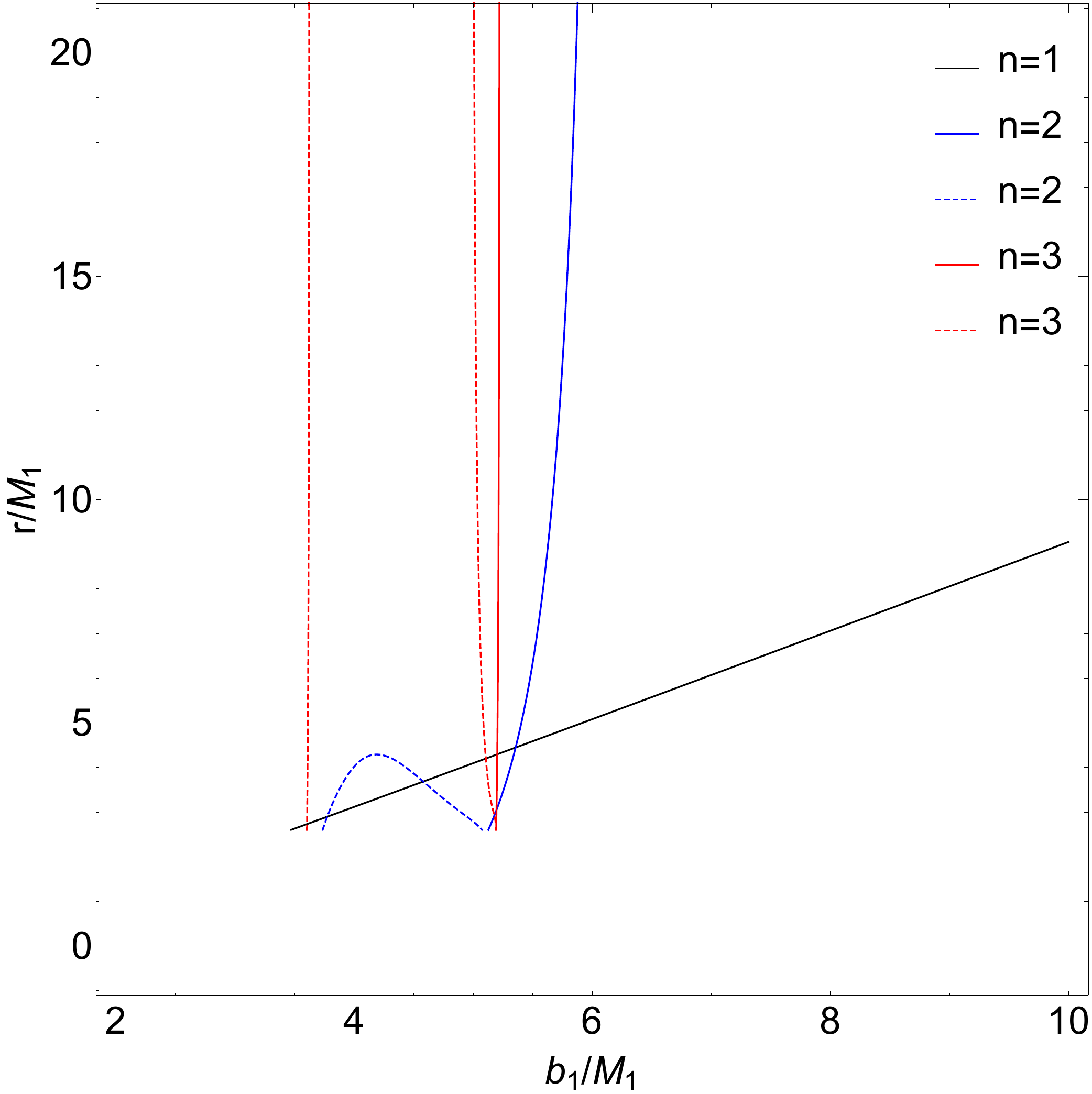}
		\end{minipage}%
	}%
	\subfigure[$\gamma=0.2$\label{transferIII}]{
		\begin{minipage}[t]{0.33\linewidth}
			\centering
			\includegraphics[width=4cm,height=4cm]{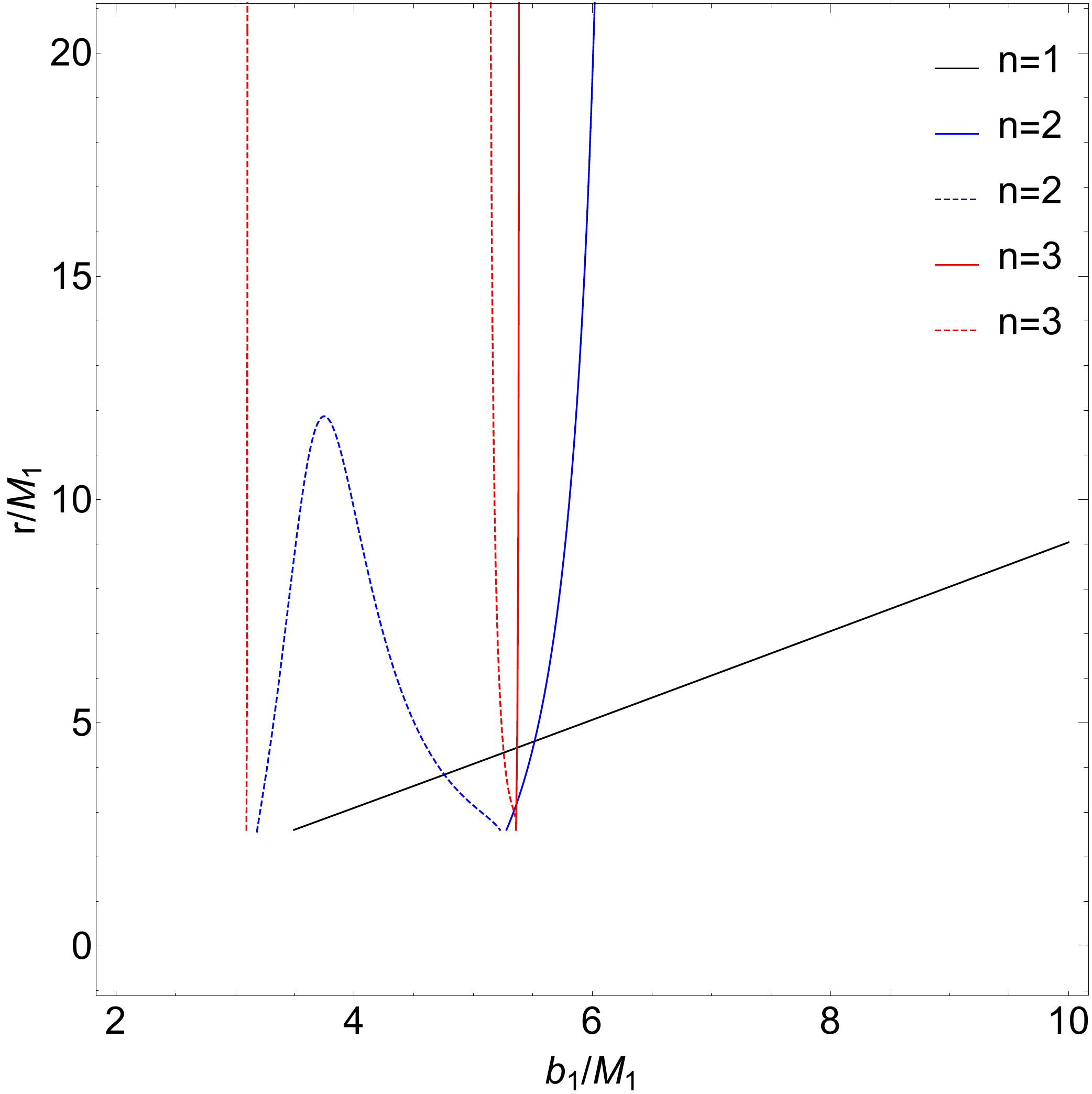}
		\end{minipage}
	}%
	\caption{The first (black lines), second (blue lines), and third (red lines) transfer functions of the ATW. Note that all the dashed lines indicate the new transfer functions of the ATW. The left, middle, and right panels correspond to $\gamma=-0.2$, $\gamma=0$, and $\gamma=0.2$, respectively. We set $M_1=1, M_2=1.2$, and $R=2.6$.}
	\label{transfer2}
\end{figure}

\subsection{\textbf{Observational appearance of the asymmetric thin-shell wormhole with two emission models of the thin accretion disk}}
\label{sec:4-3}

Now, we investigate the observational appearance of the ATW by considering two specific emission models of the thin accretion disk. The emission from the thin accretion disk can be approximated by a Gaussian function~\cite{Zhang:2013fwa}. Here, the mass parameter $M_{1}$ is still assumed to be $1$. The innermost stable circular orbit is labeled as $r_{isco}$. Next, we consider two typical emission models of the thin accretion disk to study the observational appearance of the ATW. 

In emission model I, the radiation function is given by
\begin{equation}\label{RDF1}
	I_1^{\mathrm{em}}(r)= \begin{cases}0 & r<r_{i s c o}, \\ \left(\frac{1}{r-\left(r_{isco}-1\right)}\right)^2 & r \geq r_{isco},\end{cases}
\end{equation}
where $r_{isco}$ denotes the inner edge of the accretion disk. Therefore, no radiation is emitted within the region smaller than the inner edge. The radiation function~(\ref{RDF1}) is ploted in Fig.~\ref{model1}~. With the radiation function, we plot the observed intensity, density plot, and local density plot of the ATW in the upper panel of Fig.~\ref{densityplot1}~. To compare with a BH with the same mass parameter and radiation function, we also plot the observational appearance of the BH in the lower panel of Fig.~\ref{densityplot1}~. From Figs.~\ref{Iobs1wh1} and~\ref{Iobs1bh1}~, we find that the spatial separations between the direct emission, lensing band and photon rings are distinct. For the observed intensity of the ATW [see Fig.~\ref{Iobs1wh1}~], the direct emission appears near the critical curve $b_1 \simeq 6.60M_{1}$ with an initial intensity of $0.42$, followed by a subsequent decrease. The range of the lensing band is confined within a narrow interval spanning from the critical curve $b_1 \simeq 5.30M_{1}$ to the critical curve $b_1 \simeq 5.85M_{1}$. The photon rings manifest sequentially near the critical curves $b_1 \simeq 3.91 M_{1}$, $b_1 \simeq 4.85 M_{1}$, and $b_1 \simeq 5.03M_{1}$. It is noteworthy that the ATW shadow exhibits two additional photon rings (near the critical curves $b_1 \simeq 3.91 M_{1}$ and $b_1 \simeq 4.85 M_{1}$) in comparison to the BH case [see Figs.~\ref{Iobs1wh1} and~\ref{Iobs1bh1}~]. Additionally, the density plot [see Fig.~\ref{densityplot1wh1}~] and the local density plot [see Fig.~\ref{densityplot1whlocal1}~] of the ATW show that the direct emission originates at the periphery of the black disk, while the narrow lensing band is confined within the black disk. Comparing to the BH case shown in Figs.~\ref{densityplot1bh1} and~\ref{densityplot1bhlocal1}~, the two additional photon rings appear closer to the center of the black disk in the shadow of the ATW.

\begin{figure}[!htbp]
	\centering
	\subfigure[Emission model I\label{model1}]{
		\begin{minipage}[t]{0.5\linewidth}
			\centering
			\includegraphics[width=6cm,height=6cm]{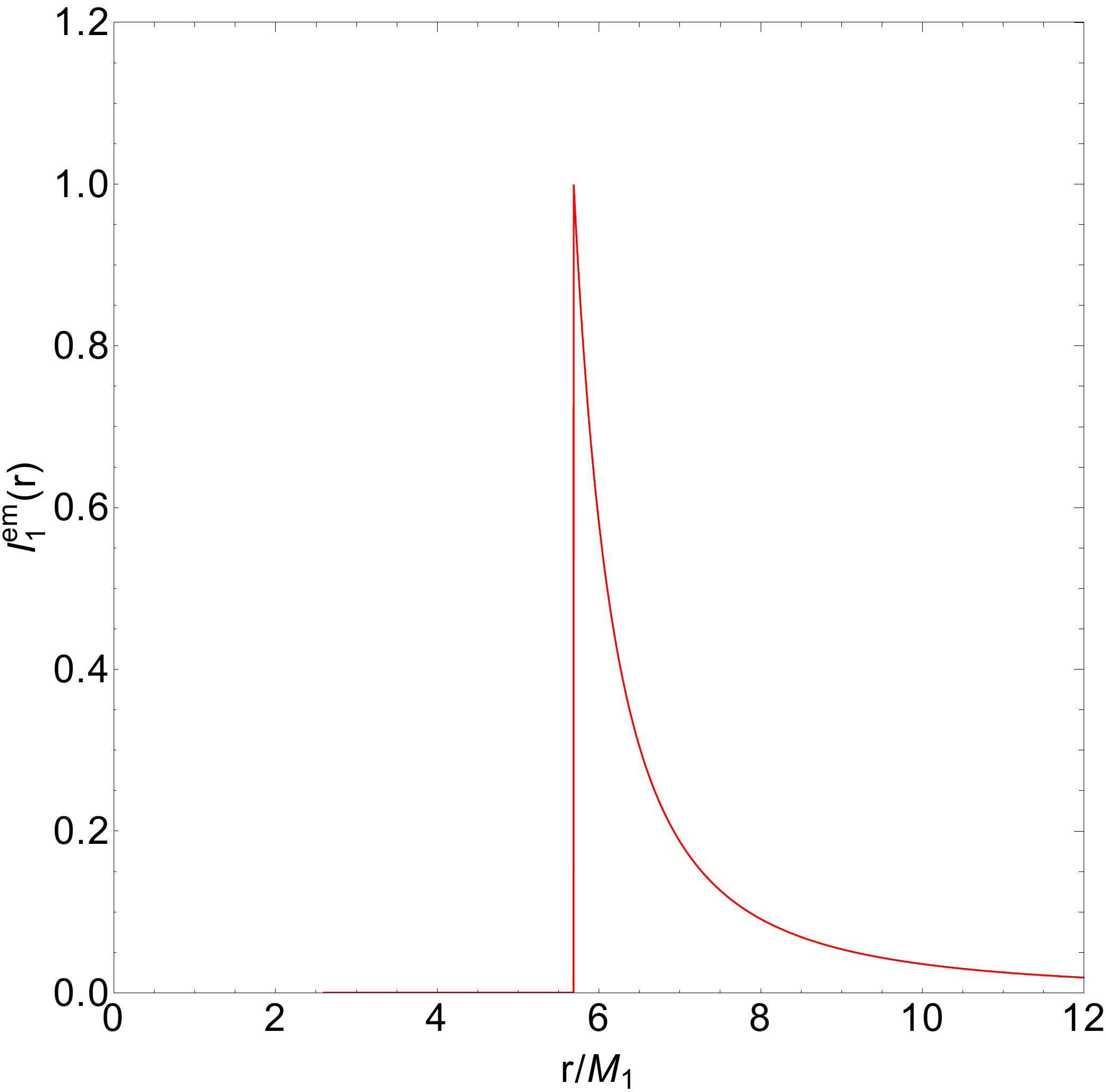}
		\end{minipage}%
	}%
	\subfigure[Emission model II\label{model2}]{
		\begin{minipage}[t]{0.5\linewidth}
			\centering
			\includegraphics[width=6cm,height=6cm]{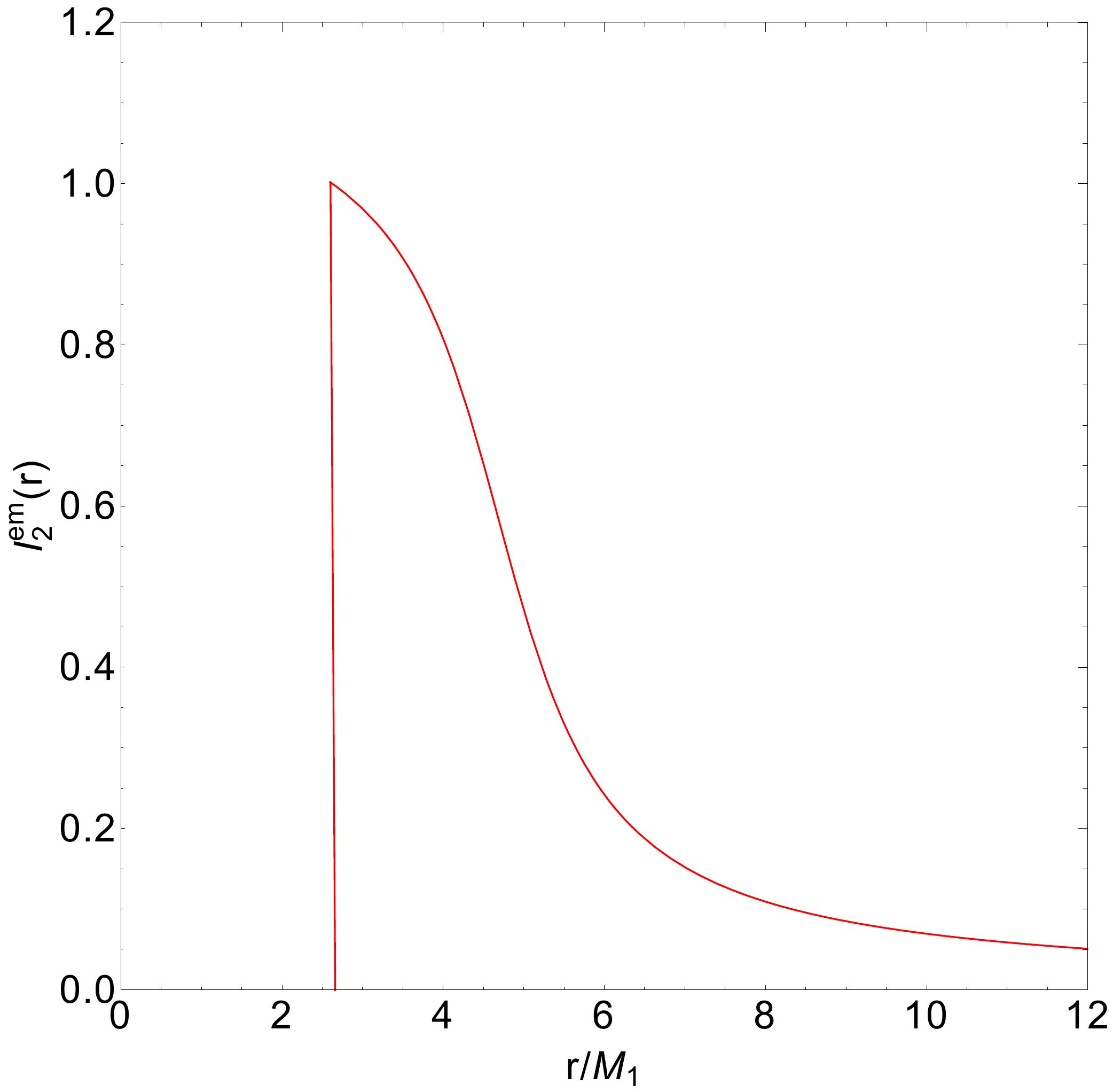}
		\end{minipage}%
	}
	\caption{The radiation fuctions of emission models I and II for the thin accretion disk.}
	\label{model}
\end{figure}

\begin{figure}[!ht]
	\centering
	\subfigure[\label{Iobs1wh1}]{
		\begin{minipage}[t]{0.33\linewidth}
			\centering
			\includegraphics[width=4cm,height=4cm]{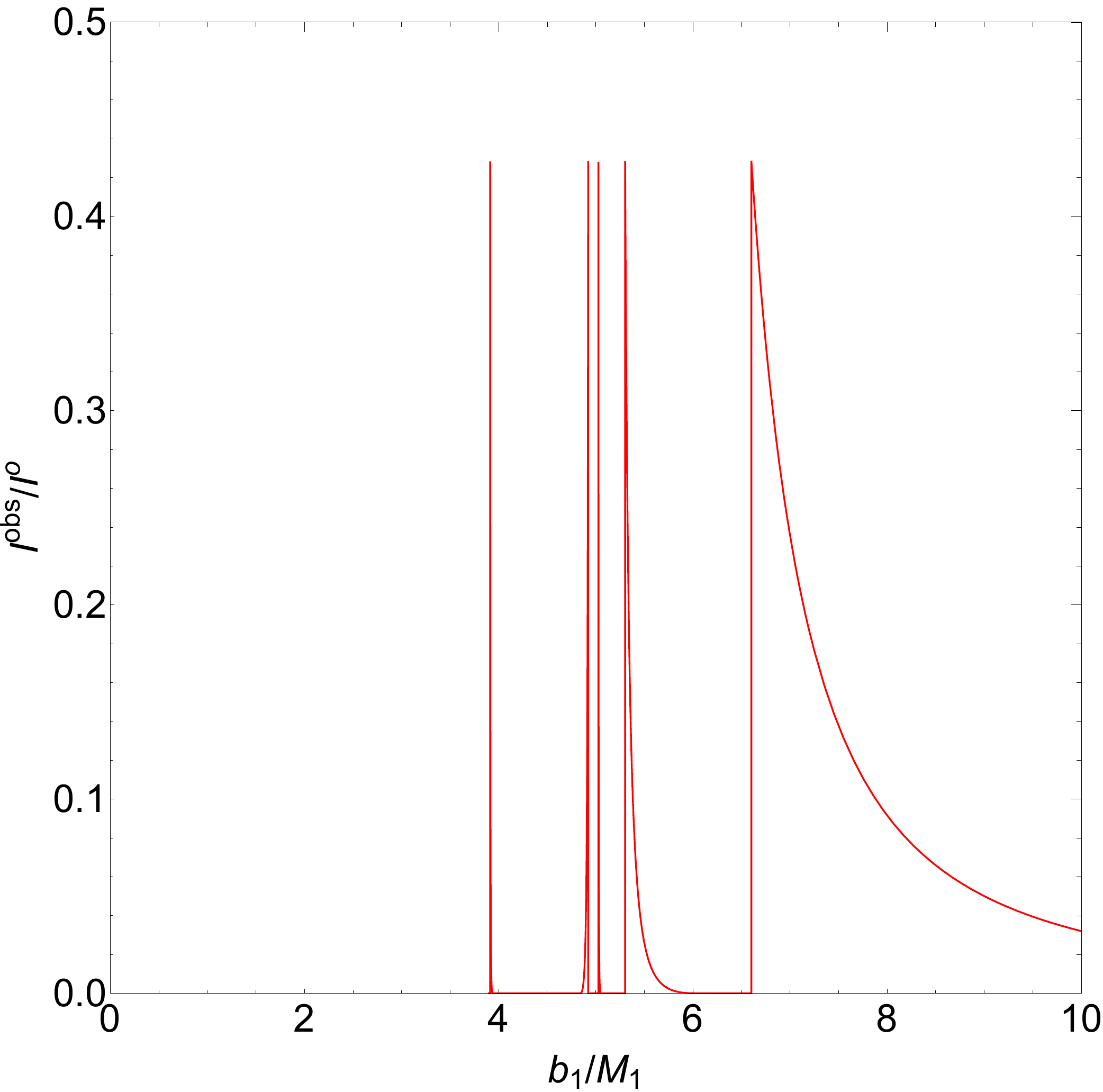}
		\end{minipage}%
	}%
	\subfigure[\label{densityplot1wh1}]{
		\begin{minipage}[t]{0.33\linewidth}
			\centering
			\includegraphics[width=4cm,height=4cm]{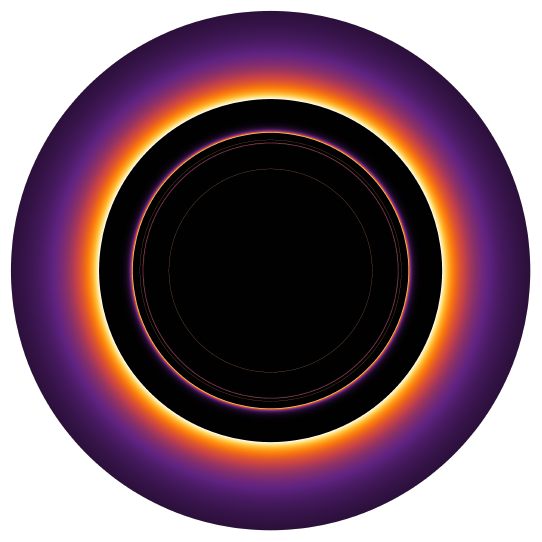}
		\end{minipage}%
	}%
	\subfigure[\label{densityplot1whlocal1}]{
		\begin{minipage}[t]{0.33\linewidth}
			\centering
			\includegraphics[width=4cm,height=4cm]{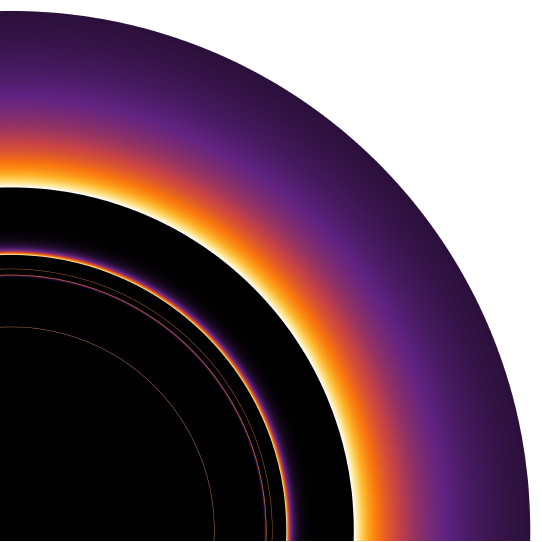}
		\end{minipage}
	}%
	\quad
	\subfigure[\label{Iobs1bh1}]{
		\begin{minipage}[t]{0.32\linewidth}
			\centering
			\includegraphics[width=4cm,height=4cm]{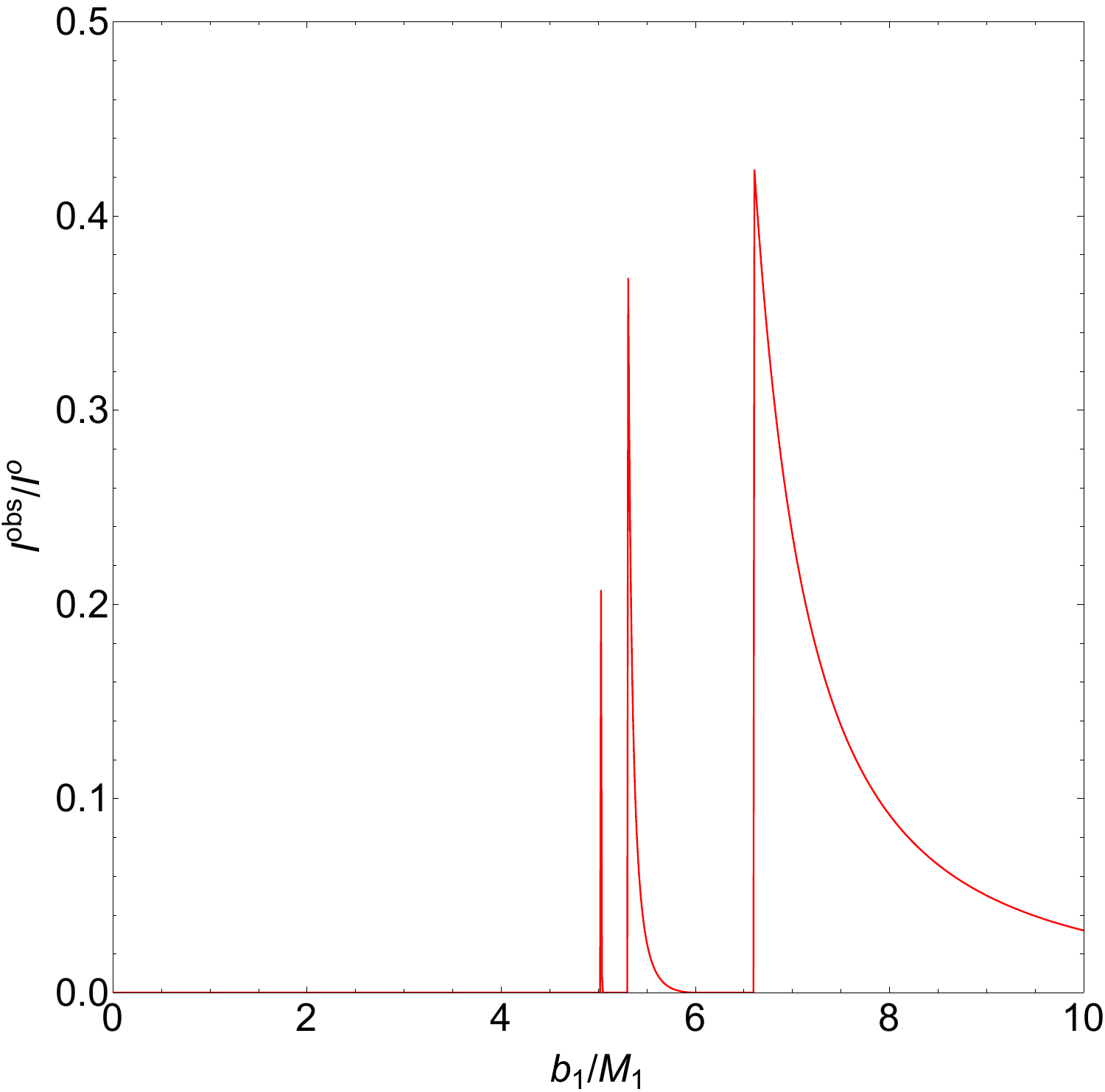}
		\end{minipage}
	}%
	\subfigure[\label{densityplot1bh1}]{
		\begin{minipage}[t]{0.32\linewidth}
			\centering
			\includegraphics[width=4cm,height=4cm]{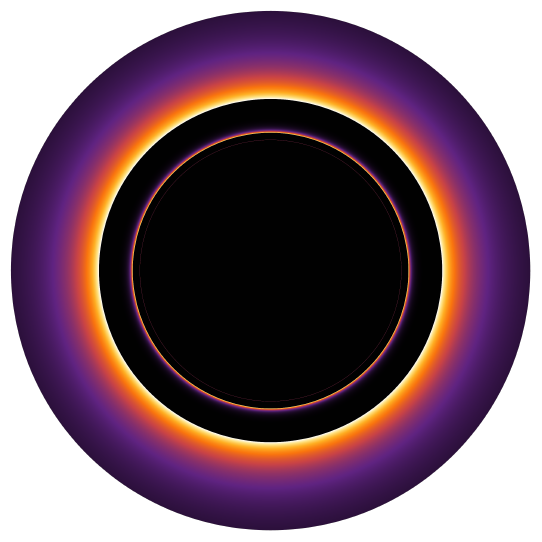}
		\end{minipage}
	}%
	\subfigure[\label{densityplot1bhlocal1}]{
		\begin{minipage}[t]{0.32\linewidth}
			\centering
			\includegraphics[width=4cm,height=4cm]{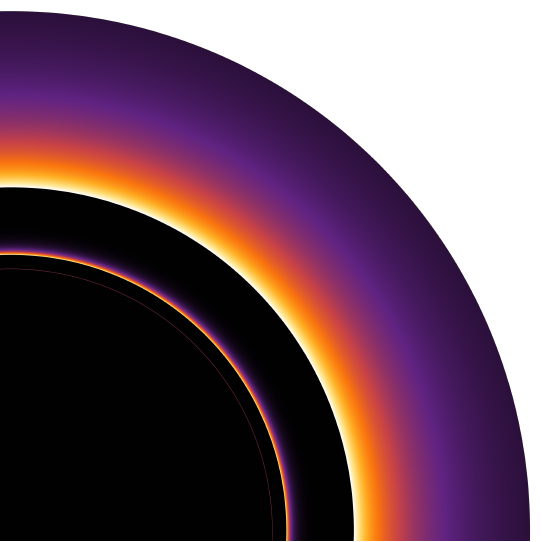}
		\end{minipage}
	}%
	\caption{\textbf{Emission Model I}-- Observed intensities (the left panel), density plots (the middle panel), and local density plots (the right panel) of the ATW (the upper panel) and the BH (the lower panel). We set $\gamma=-0.2$, $M_1=1$, $M_2=1.2$, and $R=2.6$.}
	\label{densityplot1}
\end{figure}

In emission model II, the radiation function is expressed as  
\begin{equation}\label{rdc2}
	I_2^{\mathrm{em}}(r)= \begin{cases}0 & r<r_{h}, \\ \frac{\frac{\pi}{2}-\tan ^{-1}\left(r-\left(r_{isco}-1\right)\right)}{\frac{\pi}{2}-\tan ^{-1}\left(r_{p h}\right)} & r \geq r_{h} ,\end{cases}
\end{equation}
where $r_{h}$ is the event horizon radius. Therefore, no radiation is emitted within the region smaller than the event horizon radius. The radiation function~(\ref{rdc2}) is plotted in Fig.~\ref{model2}~. From Fig.~\ref{model}~, it is evident that emission model II has a more gradual decrease in the radiation function compared to emission model I. With the radiation function, we plot the observed intensity, density plot, and local density plot of the ATW in the upper panel of Fig.~\ref{densityplot2}~. From Fig.~\ref{Iobs2wh1}~, there is a overlap between the areas of the direct emission, lensing band, and photon rings of the ATW. The direct emission initiates at approximately $b_1 \simeq 2.76M_{1}$, while the photon rings are encompassed within the lensing band, giving rise to a luminous ring structure with multiple layers [see Figs.~\ref{densityplot2wh1} and \ref{densityplot2whlocal1}~]. For a BH with the same mass parameter and radiation function, we plot the observational appearance of the BH in the lower panel of Fig.~\ref{densityplot2}~. 

Comparing emission models I and II, we find that there exists an additional lensing band between the critical curves $Z b_{ c_{2}} \simeq 3.89$ and $b_{ c_{1}} \simeq 5.01$ in emission model II. The result implies that the new second transfer function for emission model II contributes to the observed intensity of the ATW.

\begin{figure}[!ht]
	\centering
	\subfigure[\label{Iobs2wh1}]{
		\begin{minipage}[t]{0.33\linewidth}
			\centering
			\includegraphics[width=4cm,height=4cm]{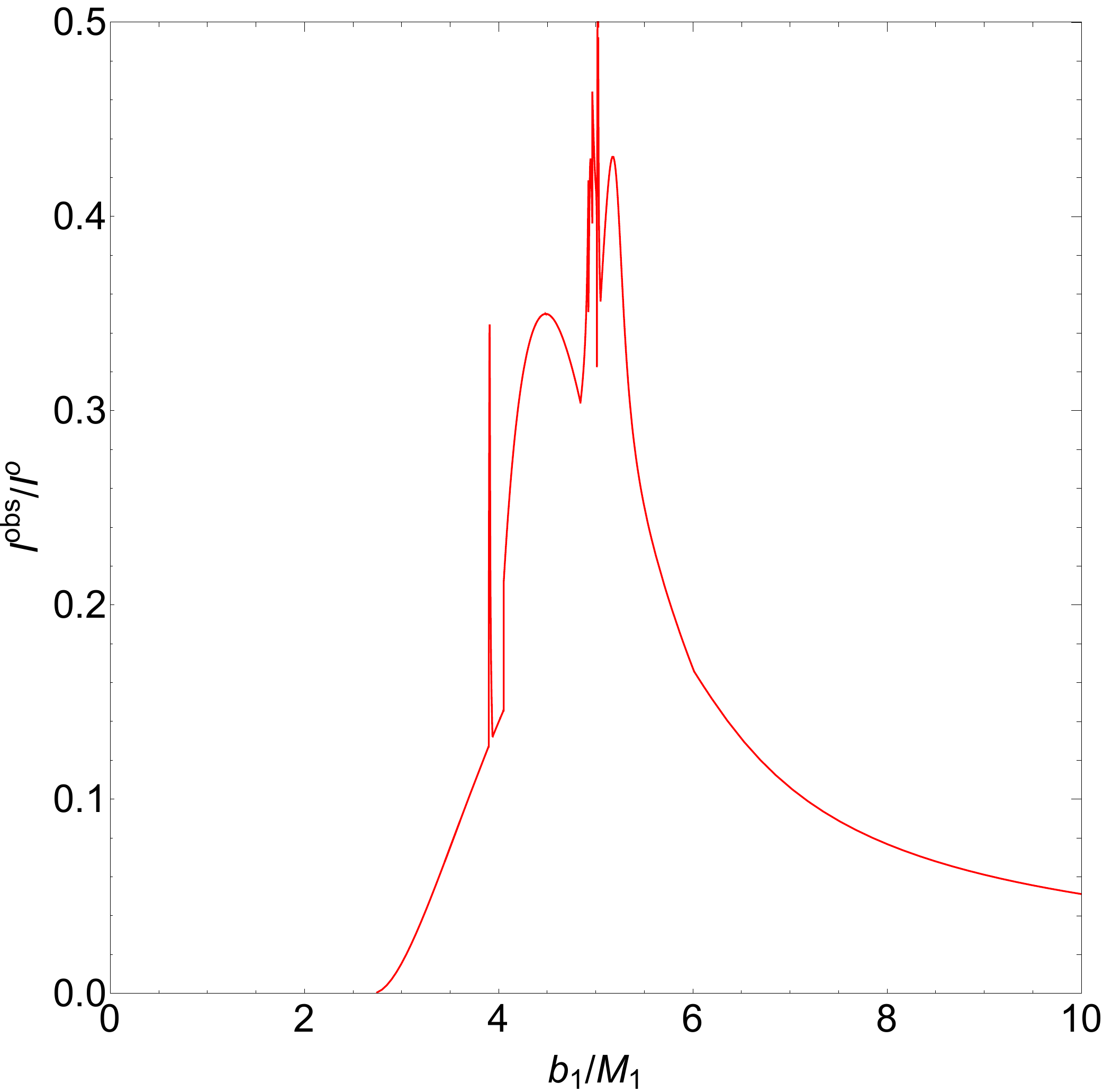}
		\end{minipage}%
	}%
	\subfigure[\label{densityplot2wh1}]{
		\begin{minipage}[t]{0.33\linewidth}
			\centering
			\includegraphics[width=4cm,height=4cm]{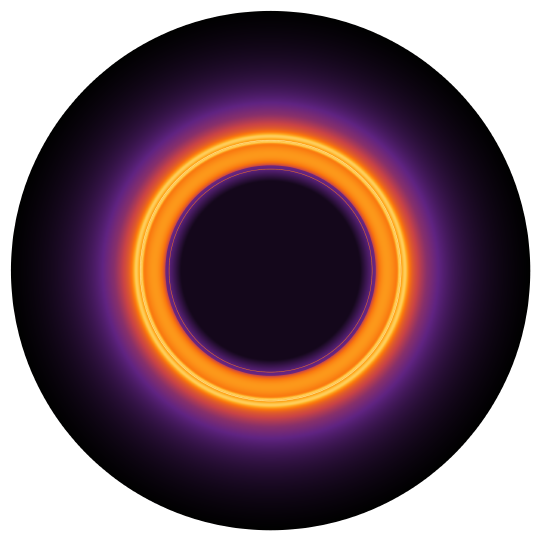}
		\end{minipage}%
	}%
	\subfigure[\label{densityplot2whlocal1}]{
		\begin{minipage}[t]{0.33\linewidth}
			\centering
			\includegraphics[width=4cm,height=4cm]{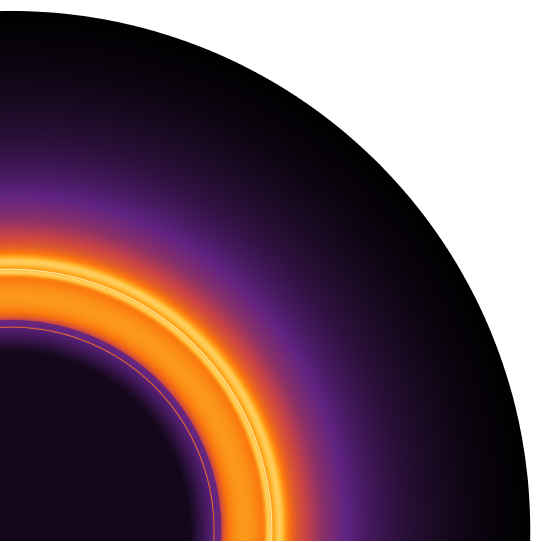}
		\end{minipage}
	}%
	\quad
	\subfigure[\label{Iobs2bh1}]{
		\begin{minipage}[t]{0.32\linewidth}
			\centering
			\includegraphics[width=4cm,height=4cm]{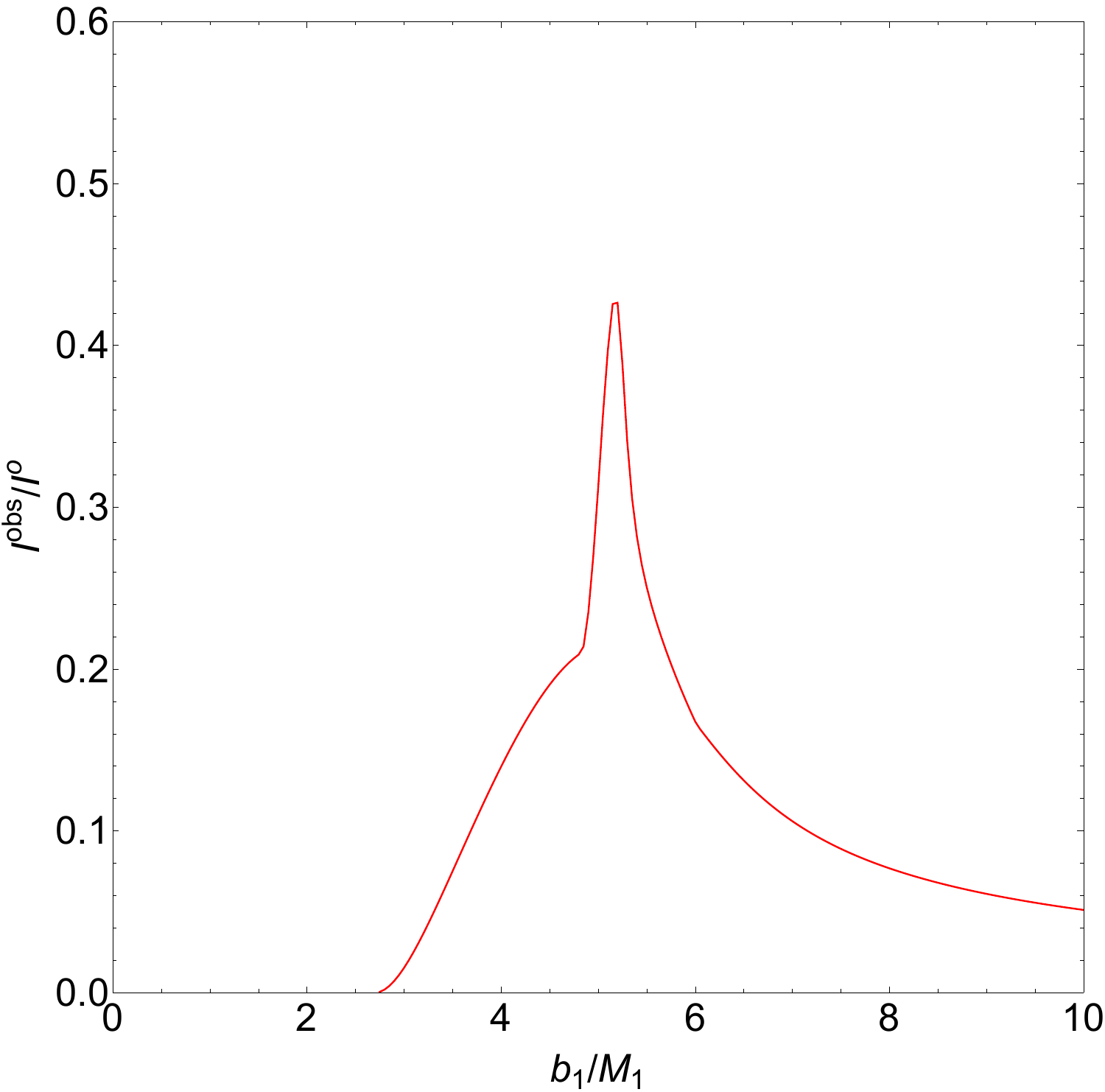}
		\end{minipage}
	}%
	\subfigure[\label{densityplot2bh1}]{
		\begin{minipage}[t]{0.32\linewidth}
			\centering
			\includegraphics[width=4cm,height=4cm]{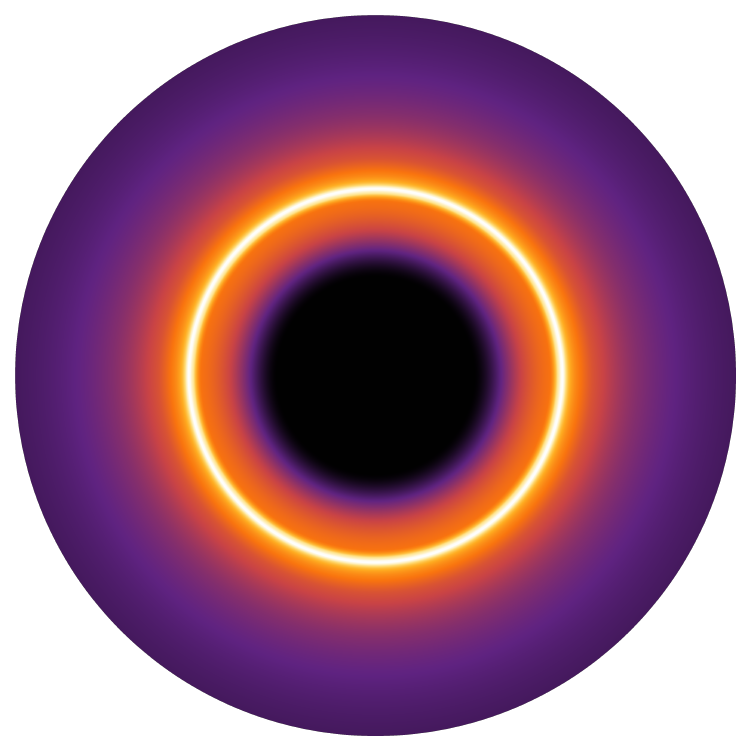}
		\end{minipage}
	}%
	\subfigure[\label{densityplot2bhlocal1}]{
		\begin{minipage}[t]{0.32\linewidth}
			\centering
			\includegraphics[width=4cm,height=4cm]{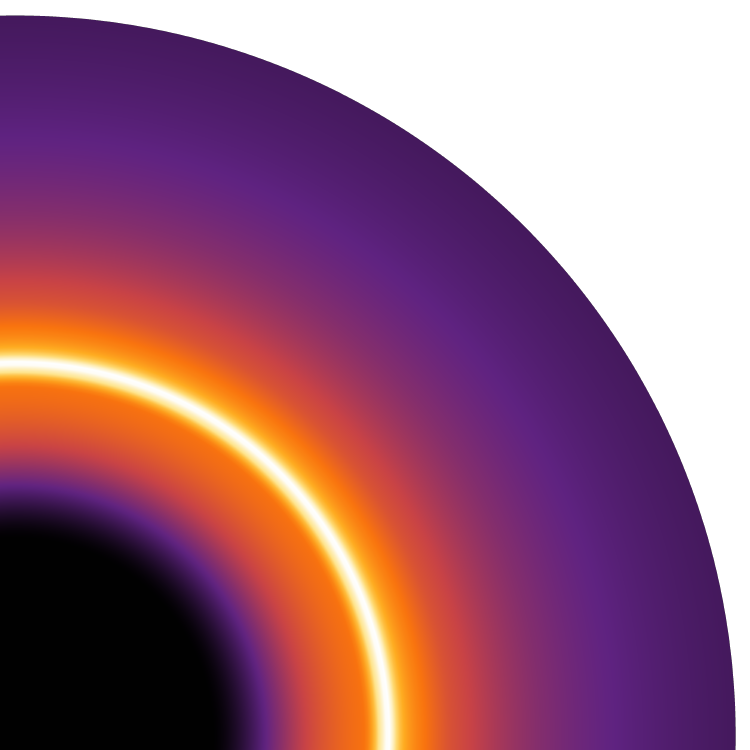}
		\end{minipage}
	}%
	\caption{\textbf{Emission Model II}-- Observed intensities (the left panel), density plots (the middle panel) and local density plots (the right panel) of the ATW (the upper panel) and the BH (the lower panel). We set $\gamma=-0.2$, $M_1=1, M_2=1.2$, and $R=2.6$.}
	\label{densityplot2}
\end{figure}

\section{Conclusions}\label{sec:5}

In this paper, we study and compare the observational appearances of the ATW and a BH with the same mass parameter in Horndeski theory. The ATW is constructed by connecting two spacetimes, $\mathcal{M}_{1}$ and $\mathcal{M}_{2}$, with different mass parameters through a throat. The observer is located in the spacetime $\mathcal{M}_{1}$, which has a smaller mass parameter. We calculate the critical impact parameter of a photon and the radius of the photon sphere in the ATW spacetime, which are listed in Table~\ref{TAB} with different values of the parameter $\gamma$. Then, we plot the effective potentials of the ATW and a BH with the same mass parameter in Fig.~\ref{potential1}~. The results indicate that when the impact parameter $b_1$ of the photon satisfies $Zb_{c_2}<b_1<b_{c_1}$, the spacetime $\mathcal{M}_{2}$ will reflect the photon back to the spacetime $\mathcal{M}_{1}$. Furthermore, we analyze the deflection angle of the photon in the ATW spacetime and plot the light trajectory for different impact parameters and values of the parameter $\gamma$ in Fig.~\ref{trajectory}~. We find that an increase in the parameter $\gamma$ will result in the expansion of the photon sphere in the ATW spacetime.

To study the observational appearance of the ATW in Horndeski theory, the orbit number of the photon is calculated. The results show that the range of the impact parameter $b_1$ will expand as $\gamma$ increases, which is consistent with the results shown in Table~\ref{TAB} and Fig.~\ref{potential1}~. We calculate the transfer functions of the ATW and find the presence of new second and third transfer functions. The new second transfer function corresponds to the ``lensing band", while the new third transfer function corresponds to the ``photon ring group". The reason for the appearance of new transfer functions is that in the ATW spacetime, when $Zb_{c_2}<b_1<b_{c_1}$, the spacetime $\mathcal{M}_{2}$ has the ability to reflect the photon back to the spacetime $\mathcal{M}_{1}$.

Finally, we consider two emission models of the thin accretion disk to study the observational appearance of the ATW in Horndeski theory. In the scenario of emission model I, as depicted in Fig.~\ref{densityplot1}~, there are two additional photon rings (compared to the BH case) positioned close to the critical curves $b_1 \simeq 3.91 M_{1}$ and $b_1 \simeq 4.85 M_{1}$, respectively. In the scenario of emission model II, we find an additional lensing band (compared to the BH case) between the critical curves $Z b_{ c_{2}} \simeq 3.89$ and $b_{ c_{1}} \simeq 5.01$. These results provide significant information for distinguishing between ATWs and BHs in Horndeski theory in terms of their observational appearances. In the future, by studying the direct emissions, lensing bands, and photon rings of UCOs, we may be able to identify the ATW and constrain the parameter $\gamma$ in Horndeski theory.

\vspace{10pt}

\noindent {\bf Acknowledgments}

\noindent
This work is supported by the National Natural Science Foundation of China (Grant No. 12147102), the Natural Science Foundation of Chongqing (Grant No. CSTB2023NSCQ-MSX0103).

\end{document}